\documentclass[
preprint,
superscriptaddress,
nofootinbib,
 amsmath,amssymb,
 aps,
prb,
floatfix,
]{revtex4-2}

\usepackage{comment}
\usepackage{graphicx}
\usepackage{dcolumn}
\usepackage{bm}
\usepackage{amsmath}
\usepackage{array}
\usepackage{longtable} 
\usepackage[dvipsnames]{xcolor}
\usepackage{ulem}

\begin{document}

\preprint{APS/123-QED}

\title{Spatially Heterogeneous Relaxational Dynamics and the Evolution of Recoverable Strain following Flow Cessation of a Ductile Nanocolloidal Glass.}

\author{Chloe W. Lindeman}
\thanks{These authors contributed equally to this work.}
\affiliation{Department of Physics and Astronomy, Johns Hopkins University, Baltimore, Maryland 21218, USA}
\author{James J. Griebler}
\thanks{These authors contributed equally to this work.}
\affiliation{Department of Chemical and Biomolecular Engineering, University of Illinois Urbana-Champaign, Champaign, IL 61801, USA}
\author{Penelope Grace Kovakas}
\affiliation{Department of Chemical and Biomolecular Engineering, University of Illinois Urbana-Champaign, Champaign, IL 61801, USA}
\author{Miaoqi Chu}
\affiliation{X-Ray Science Division, Argonne National Laboratory, Argonne, Illinois 60439, USA}
\author{Qingteng Zhang}
\affiliation{X-Ray Science Division, Argonne National Laboratory, Argonne, Illinois 60439, USA}
\author{Suresh Narayanan}
\affiliation{X-Ray Science Division, Argonne National Laboratory, Argonne, Illinois 60439, USA}
\author{James L. Harden}
\affiliation{Department of Physics, University of Ottawa, Ottawa, Ontario K1N 6N5, Canada}
\author{Simon A. Rogers}
\affiliation{Department of Chemical and Biomolecular Engineering, University of Illinois Urbana-Champaign, Champaign, IL 61801, USA}
\author{Robert L. Leheny}
\affiliation{Department of Physics and Astronomy, Johns Hopkins University, Baltimore, Maryland 21218, USA}

\date{\today}%
 
\begin{abstract}
We report a combined rheology and x-ray photon correlation spectroscopy (XPCS) study of the structural and mechanical relaxation of a ductile, nanocolloidal glass following the cessation of shear flow. After the glass is sheared to 300\% strain at various shear rates and then held at fixed strain, the stress undergoes a protracted, quasi-logarithmic decay with hold time that depends weakly on the initial strain rate. Recovery rheology measurements reveal that this stress relaxation is accompanied by a logarithmic decrease in the elastic component of the recoverable strain; hence, the rates of decrease of the stress and recoverable strain are proportional. XPCS measurements during the stress relaxation reveal dynamics dominated by a convection-like backflow that is divided into two dynamically distinct regions indicative of banded motion.  In one region, the flow can be modeled by an affine strain, while in the other region the glass moves as a plug while undergoing slow, glassy relaxation. The rates of these dynamics approximately track the rate of loss of recoverable strain, indicating this motion is the predominant microscopic mechanism driving the conversion of recoverable to unrecoverable strain during stress relaxation. In contrast, XPCS measurements during strain recovery reveal purely affine flow with no evidence of heterogeneity and with strain rates that agree quantitatively with the rheometry measurements. Together, these results provide a unified microscopic picture connecting the evolving internal dynamics of a ductile glass to its macroscopic mechanical relaxation following flow cessation.
\end{abstract}

\maketitle

\section{Introduction}
\label{sec:intro}

Many soft disordered materials, including colloidal gels, foams, emulsions, and concentrated suspensions, behave as yield stress fluids~\cite{BonnRMP2017}.  A yield stress fluid is defined by its ability to respond to applied stress as a solid below some critical stress and to flow as a liquid above it.  Such materials are ubiquitous in industrial and biological systems, appearing in applications ranging from food processing and pharmaceuticals to oil recovery and additive manufacturing. Understanding what happens when flow is suddenly stopped is important in these contexts, as the transient structural recovery of the material directly governs practical outcomes such as the ability of a ceramic piece to hold its shape during additive manufacturing or of a gel to re-solidify after injection. The cessation of flow can potentially trigger a complex competition among viscous relaxation, elastic recoil, and thixotropic restructuring, making it a rich problem for investigating the interplay between microstructural dynamics and macroscopic mechanical response. This behavior further dictates the extent to which the material maintains a memory of the preceding flow~\cite{MohanPRL2013, VinuthaPNAS2024, PadamataPRL2026, VasishtSM2022}, which can have  consequences for the material's long-term properties and for strategies to erase such memory~\cite{ChoiRheoActa2020,EderaPRX2025}.

Despite the significance of the behavior of yield stress fluids following flow cessation, many questions remain, especially about the microscopic processes governing the behavior.  In this paper, we report experiments on a model yield stress fluid -- a ductile, nanocolloidal glass -- that address this issue by combining rheology measurements with {\it in situ} x-ray photon correlation spectroscopy (XPCS) measurements to connect the evolving microstructural dynamics with the mechanical response.  These experiments complement a previous study by our group in which we used rheology in conjunction with XPCS (rheo-XPCS)~\cite{Leheny_COCIS_2015} to investigate stress relaxation in a nanocolloidal glass~\cite{chen2020microscopic}.  The focus of that study was to understand the microscopic origins of the protracted decay in the stress that occurs after the application of step strains to values close to the yield strain.  A significant finding of that work was the identification of slow, strain-like motion described as a ``convective backflow'' as the dominant microscopic dynamics underlying the decay in stress~\cite{chen2020microscopic}.  Recent work by Mutneja and Schweizer has shown that a theory for mechanically driven amorphous solids based on a nonlinear Langevin equation captures this convective backflow phenomenon, which results from a coupling between structural evolution and stress relaxation following step strains~\cite{MutnejaJoR2026}.  Here, we address experimentally the related question of how a nanocolloidal glass behaves following cessation of shear flows well beyond the yield strain.  We find that, as in the case of small-amplitude step strains, the mechanical response after flow cessation is characterized by a protracted decay in the stress and that the underlying microscopic dynamics are dominated by a convective backflow.  However, the dynamics are more complicated than in the case of smaller step strains and show a spatial heterogeneity that is indicative of banded motion.  To better understand the evolution of the glass following the flow cessation, we also perform recovery rheology measurements to characterize the changes in recoverable strain during the stress relaxation~\cite{lee2019structure}.  We find that the recoverable strain decreases at a rate that is proportional to the rate of decay in the stress.  The rate at which the recoverable strain decreases further correlates with the evolving microscopic dynamics observed with XPCS, thus providing a unifying picture of the changes in the glass following flow cessation.  

The remainder of the paper is organized as follows. In Sec.~\ref{sec:materials_and_methods} we describe the preparation of the nanocolloidal glass and the procedures involved in the rheo-XPCS experiments.  The results and analysis are presented in Sec.~\ref{sec:results}.  We start Sec.~\ref{sec:results} with results for the stress relaxation and evolution in recoverable strain following flow cessation, showing that the rates of change of the stress and recoverable strain are proportional.  We then present the {\it in situ} XPCS results that demonstrate the spatially heterogeneous convective backflow underlying this evolving rheology.  We finish Sec.~\ref{sec:results} by comparing the microscopic dynamics during stress relaxation with those during strain recovery.  Finally, in Sec.~\ref{sec:conclusion} we provide a discussion of the findings and some concluding remarks.  In the appendix we provide a table listing the variables introduced throughout the manuscript and their definitions.

\section{Materials and Methods}
\label{sec:materials_and_methods}

\subsection{Nanocolloidal Glass Preparation}
The glass was composed of a suspension of spherical charged silica colloids with diameters of 26 nm (Ludox TM50, Sigma Aldrich) in water. The original volume fraction of the aqueous suspension was 30\% according to the manufacturer, and the particles were stabilized by negatively charged surfaces and counterions in the suspension. 15 mL of the suspension was centrifuged at $17000g$ for 30 minutes, and the supernatant was poured off, leaving a solid plug of material.  Several plugs produced in this way were combined with a small amount of water, and a Thinky Speed Mixer was employed to homogenize the sample.  Following the experiments, a portion of the sample was weighed, then dried and reweighed, to measure the solid fraction, which corresponded to a colloidal volume fraction of 0.42, assuming a silica density of 2 g/cm$^3$.  This volume faction is in the range where Ludox suspensions behave as ductile soft solids~\cite{PhilippePRE2018}, and similar suspensions have been employed in previous rheo-XPCS studies~\cite{chen2020microscopic,DonleyPNAS2023,ChenPRM2025}.

\subsection{Rheo-XPCS}
The rheo-XPCS experiments were carried out at Sector 8-ID of the Advanced Photon Source.  The sample was contained in a Couette cell of a stress-controlled rheometer (Anton Paar MCR 702) mounted on the beam line, enabling rheological tests in parallel with x-ray scattering measurements.  A 12.4 keV, partially coherent x-ray beam was focused to an $8\times8$ $\mu$m$^2$ spot on the sample.  An area detector (EIGER2 X CdTe 4M) positioned 11.6 m after the sample measured the scattering intensity over the range of scattering wave vectors 0.04 nm$^{-1} < |\bf{q}| <$ 0.51 nm$^{-1}$.  The Couette cell was made of thin-walled polycarbonate with inner and outer diameters of 11.0 and 11.4 mm, respectively, and measurements were performed with the incident beam directed through the center of the cell so that it was parallel to the shear-gradient direction.  In the small-angle scattering regime, the scattering wave vectors were therefore in the flow-vorticity plane.  In a typical measurement, the glass was first subjected to a preshear consisting of 10 cycles of oscillatory strain to 1000\% strain at 1 rad/s followed by 60 seconds at zero applied stress to erase the mechanical history and then a brief measurement of the linear rheology to test for consistency.  (See the Supplemental Material (SM)~\cite{SMnote} for rheology results.)  The glass was then subjected to steady shear to 300\% strain at a fixed shear rate.  The strain was then held constant at 300\%, and the stress required to maintain the constant strain was monitored as a function of hold time $t_h$.  Measurements were performed for six different shear rates between $5.62\times10^{-4}$ s$^{-1}$ and 0.01 s$^{-1}$.  In the measurements of the recoverable strain, the glass was first sheared to 300\% strain at 0.01 s$^{-1}$, and the strain was held fixed at 300\% for a specified hold time.  The stress was then set to zero, and the strain as a function of recovery time $t_r$ was measured. Throughout the stress relaxation and strain recovery measurements, coherent x-ray scattering images were collected at 10 or 16.67 frames per second depending on the measurement to track the microstructural dynamics.  An example scattering image and further details about the analysis procedure are provided in the SM.  

\section{Results}
\label{sec:results}
\subsection{Stress Relaxation}
\label{subsec:stressrelaxation}
\begin{figure}
\includegraphics[width=6cm]{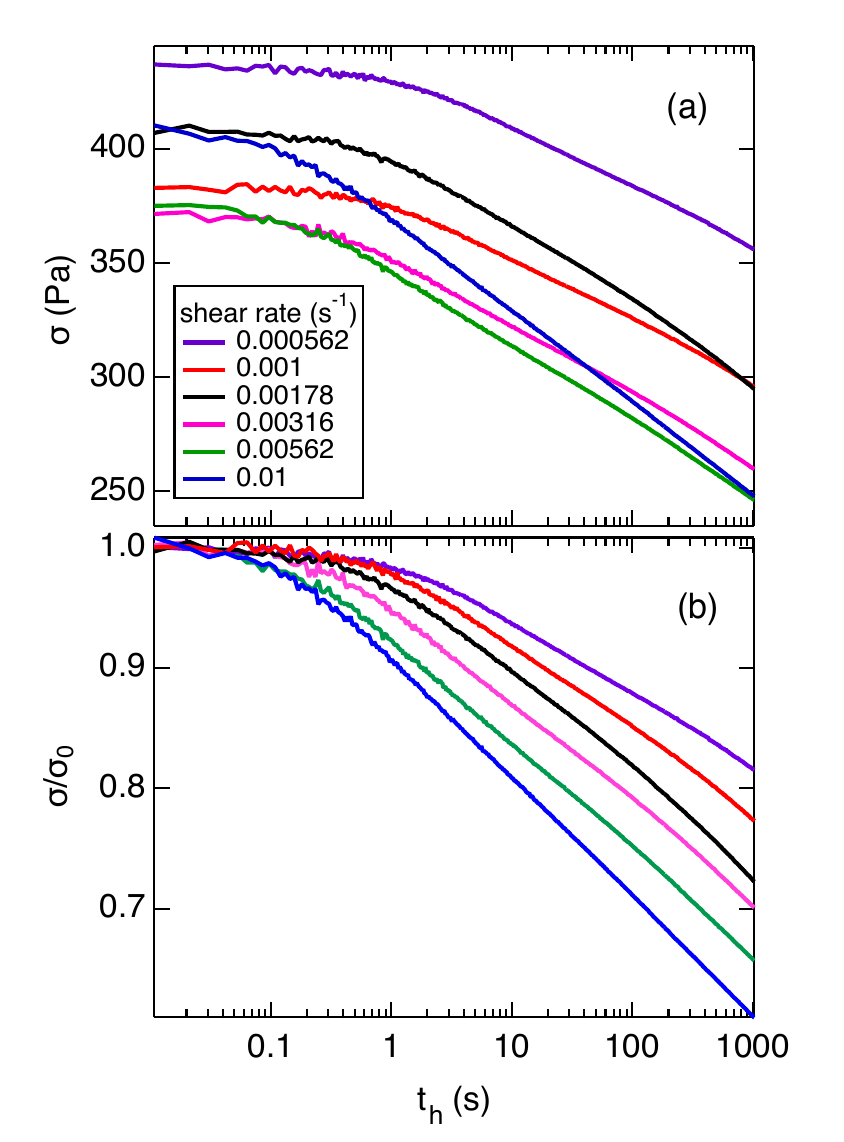}
\caption{(a) Stress as a function of hold time following cessation of shear flow to 300\% strain at various shear rates, as indicated in the legend. (b) Stress following the various shear rates normalized by the initial value at small hold times.}
\label{stressrelax}
\end{figure}

\begin{figure}
\includegraphics[width=6cm]{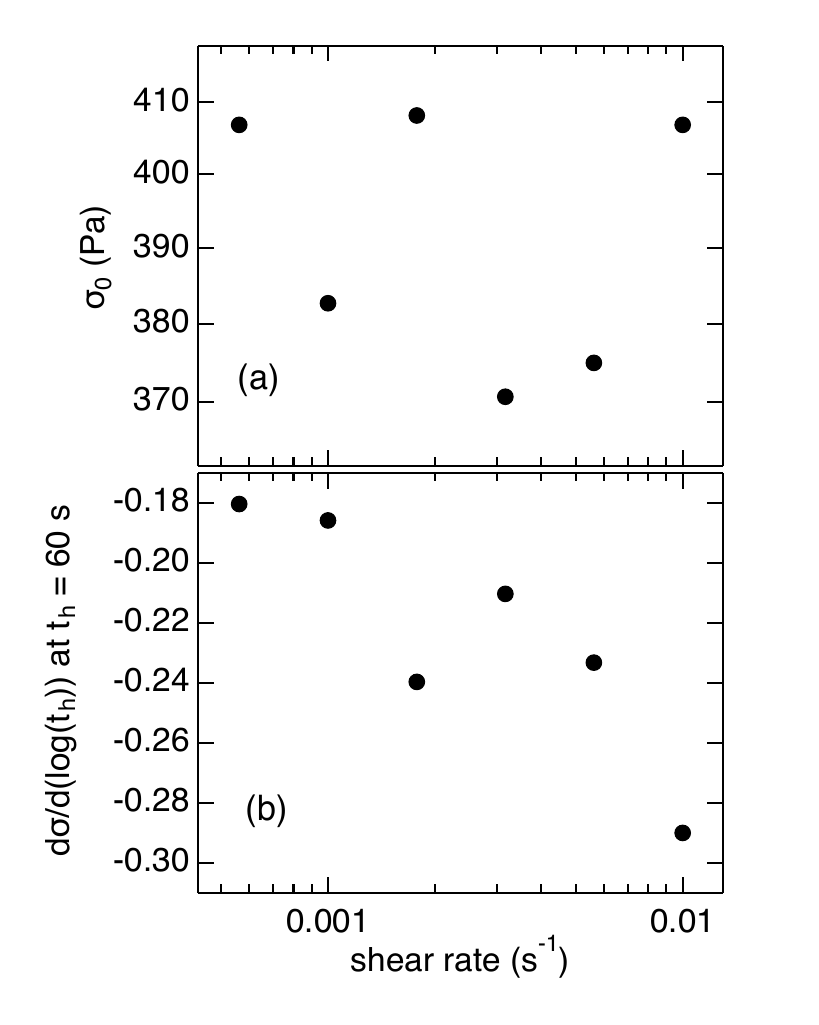}
\caption{(a) Small hold time values of the stress and (b) rate of stress relaxation (at $t_h = 60$ s) on a logarithmic time scale following cessation of shear flow to 300\% strain as a function of shear rate.}
\label{stressrelaxrate}
\end{figure}

Figure \ref{stressrelax}(a) shows the stress $\sigma$ as a function of hold time $t_h$ following the cessation of shear flow to 300\% strain at various shear rates.  After a brief initial stage ($t_h \lesssim 1$ s) during which the stress is roughly constant, the stress displays a protracted decay in which it decreases approximately logarithmically with hold time.  The small-hold-time values $\sigma_0$ following the shear flow at the different rates are shown in Fig.~\ref{stressrelaxrate}(a), and the stress normalized by these values is shown in Fig.~\ref{stressrelax}(b).  The small-hold-time values have no clear dependence on the preceding shear rate, but the rate of stress decay depends weakly on the preceding rate.  To characterize this dependence, we show in Fig.~\ref{stressrelaxrate}(b) the logarithmic decay rate, $d\sigma/d(log(t_h))$, at $t_h = 60$ s for the different shear rates.  In hard-sphere and repulsive colloidal suspensions, the stress decay following flow cessation displays different time dependencies depending on particle concentration, from exponential or stretched exponential at lower concentration to weak power-law or quasi-logarithmic at higher concentrations deep in the glass phase~\cite{McKenna_JoR_2009,Ballauff_PRL_2013,Ranjini_SM_2010,MohanJoR2015,Pamvouxoglou_JoR_2021,MutnejaJoR2026}.  These trends indicate that our nanocolloidal glass is at the concentrated end of this spectrum.  Such quasi-logarithmic stress relaxation following flow cessation has also been observed in other materials that behave as yield stress fluids, including colloidal gels~\cite{NegiPRE2009,LidonRheoActa2017,Moghimi_SM_2017}.  A key objective of the present study is to gain insight into the internal dynamics of the glass that underlie this stress relaxation.  

\subsection{Strain Recovery}
\label{subsec:strainrecovery}

\begin{figure}
\includegraphics[width=6cm]{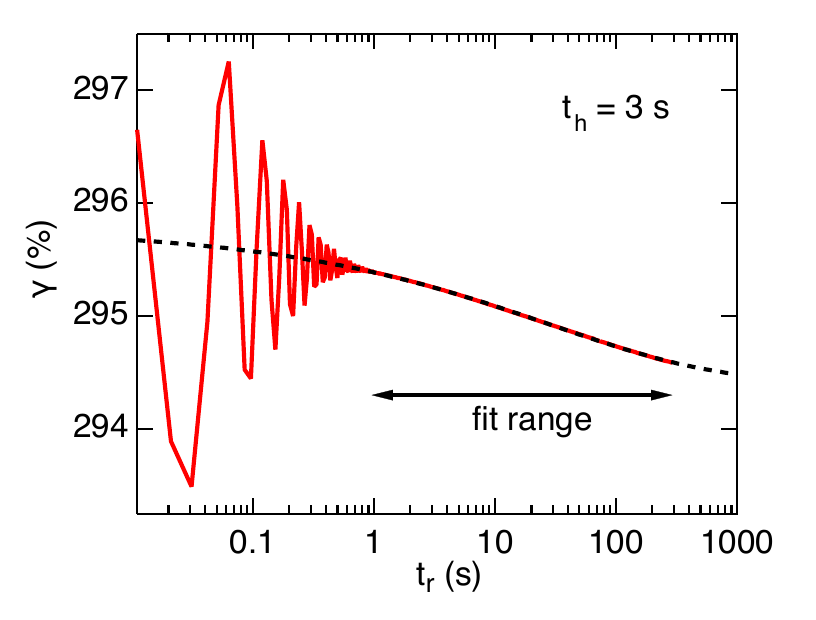}
\caption{Strain as a function of recovery time following cessation of shear flow to 300\% strain at 0.01 s$^{-1}$ and stress relaxation at fixed strain for a hold time of 3 s.  The dashed line displays the result of a stretched-exponential fit to the data over the range 1 s $< t_r <$ 290 s.}
\label{strainrec_3s}
\end{figure}

To characterize the internal changes in the glass during stress relaxation, we measured the evolution in recoverable strain following cessation of the shear flow, focusing on an initial shear rate of 0.01 s$^{-1}$.  Figure \ref{strainrec_3s} shows the strain as a function of the recovery time $t_r$ after the stress was set at zero in a measurement in which the stress was first allowed to relax with the strain fixed at 300\% for 3 s.  Once the stress was set at zero, the strain underwent an effectively discontinuous jump downward, corresponding to a sudden, elastic strain recovery. Inertial effects following the jump caused damped oscillatory motion of the Couette bob, which was reflected in the measured strain.  These oscillations died down by $t_r \approx 1$ s, after which the total strain slowly decreased (the recovered strain slowly increased) in a highly protracted manner characteristic of glassy dynamics that continued beyond the end of the measurement ($t_r \approx 290$ s).  Such a two-step strain recovery, with a rapid jump followed by a slow recovery, has been seen previously in colloidal gels~\cite{NegiPRE2009,LidonRheoActa2017} and glasses~\cite{Pamvouxoglou_JoR_2021,DonleyPNAS2023}.  Lockwood and Fielding also recently showed that the soft glassy rheology (SGR) model produces such a two-step strain recovery following flow cessation~\cite{LockwoodJoR2025}.  Interestingly, in the model the slow component of the strain recovery is identified with ``reverse'' plastic events in regions of the material that are placed in states of negative strain by the initial, step-like strain recovery.  

\begin{figure}
\includegraphics[width=6cm]{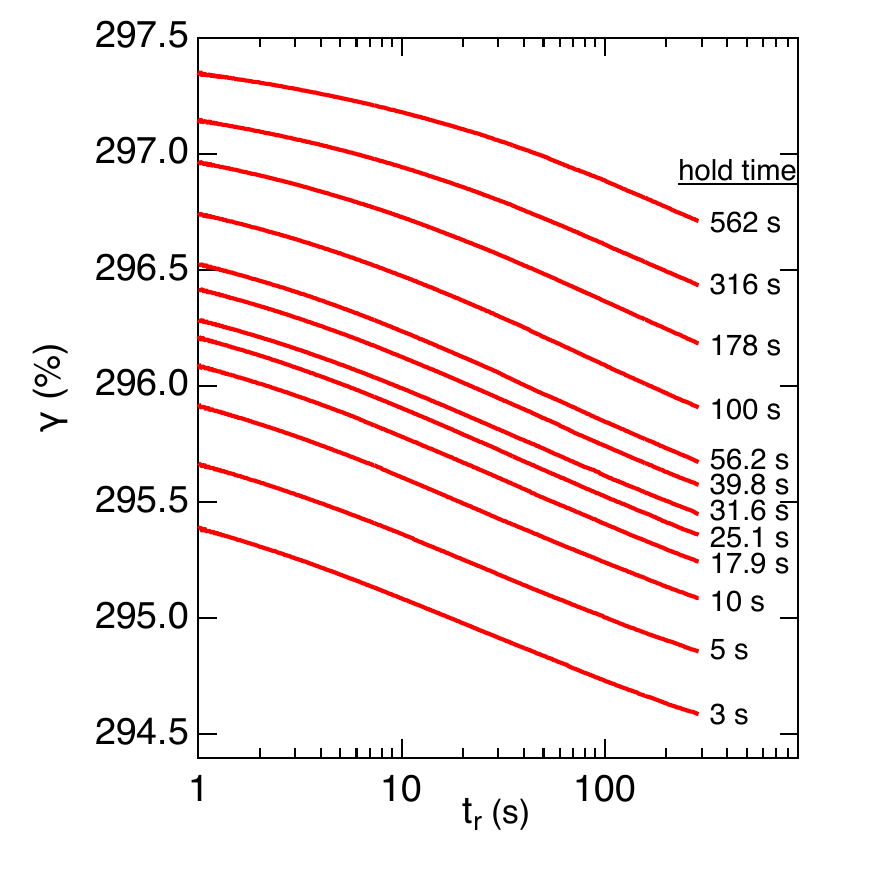}
\caption{Strain as a function of recovery time following cessation of shear flow to 300\% strain at 0.01 s$^{-1}$ and stress relaxation at fixed strain for various hold times.}
\label{strainrec_alltw}
\end{figure}

\begin{figure}
\includegraphics[width=6cm]{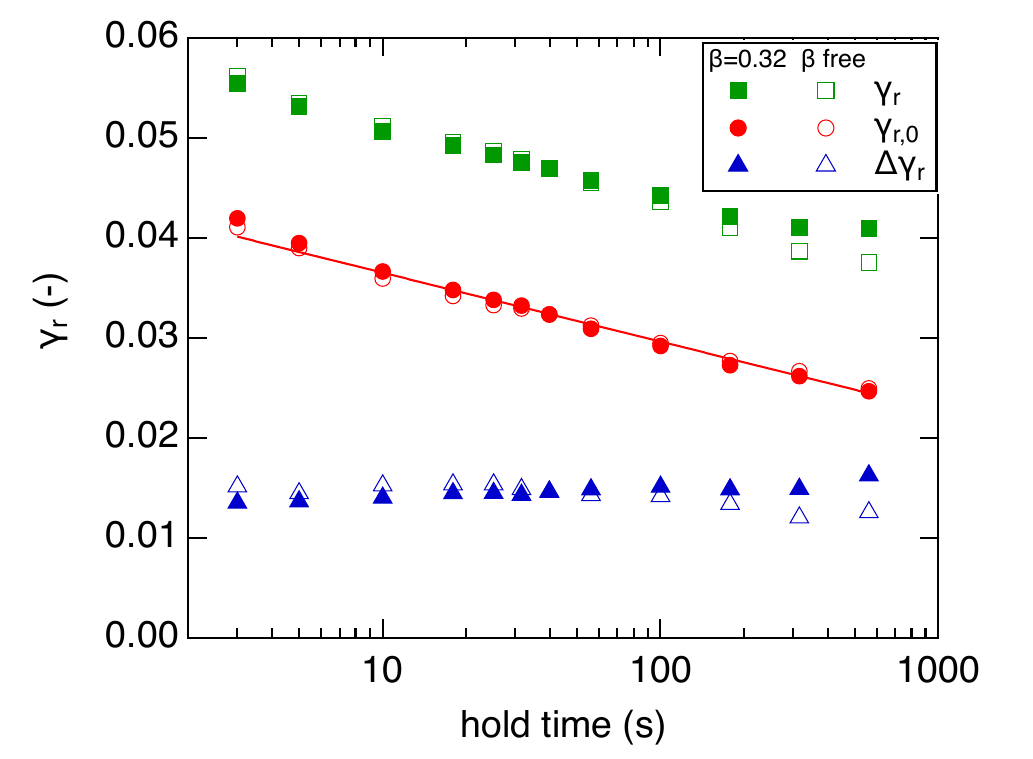}
\caption{Magnitudes of the elastic, rapidly recovered strain (red circles), the glassy, slowly recovered strain at asymptotically late $t_r$ (blue triangles), and the total recoverable strain (green squares) as functions of hold time.  Results are shown from fits to the strain recovery where the stretched-exponential shape parameter $\beta_r$ is a free parameter (open symbols) and where it is fixed to its average value of 0.32 (solid symbols).  The solid line show the result of a logarithmic fit to the elastic, rapidly recovered strain.}
\label{gamma_rec}
\end{figure}

\begin{figure}
\includegraphics[width=6cm]{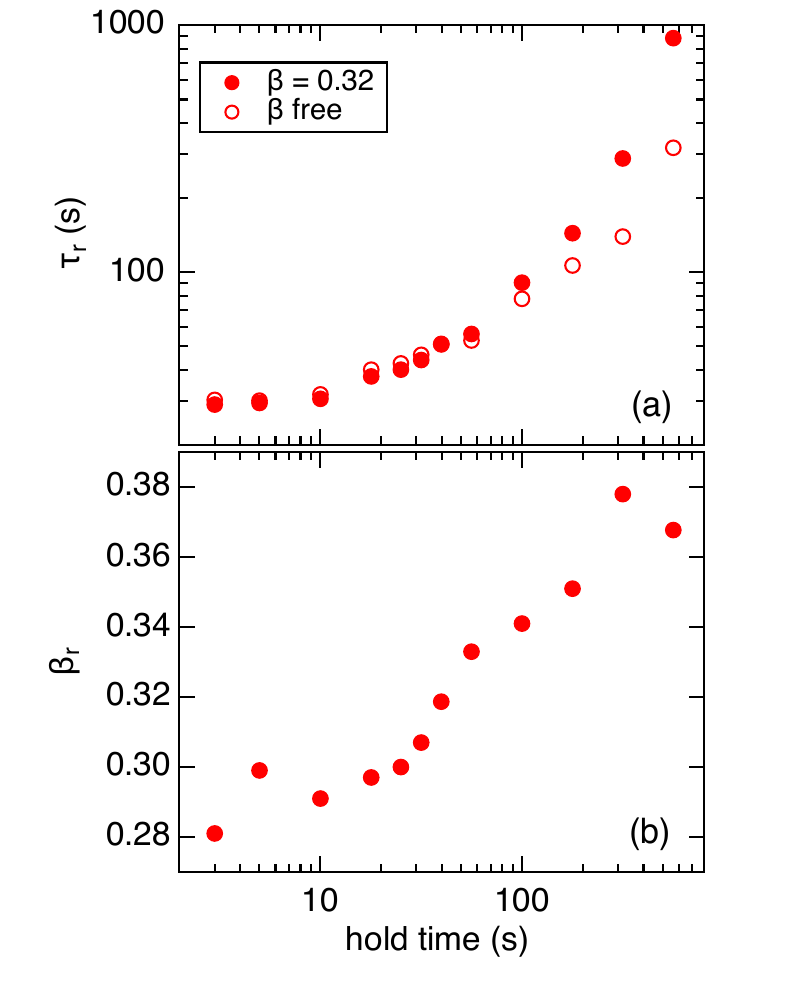}
\caption{(a) Characteristic time scale $\tau_r$ and (b) shape parameter $\beta_r$ obtained from stretched exponential fits to the strain recovery following cessation of shear flow to 300\% strain at 0.01 s$^{-1}$ and stress relaxation at fixed strain for various hold times.  The results for the characteristic time scale include those where $\beta_r$ is a free parameter in the fitting (open) and where it is fixed to its average value of 0.32 (solid).}
\label{beta_and_tau_rec}
\end{figure}

Figure \ref{strainrec_alltw} shows the strain as a function of recovery time from similar measurements in which the stress was allowed to relax at 300\% strain for various hold times $t_h$ before the stress was set at zero.  Only the strains at recovery times $t_r > 1$ s are shown in order to exclude the damped oscillations at shorter recovery times.  Following all hold times, the recoverable strain contains an instantaneous, elastic component and a slow, glassy component.  The strain recovery evolves with increasing hold time in two ways.  First, the size of the initial, elastic jump decreases with increasing $t_h$, and second, the rate of the glassy recovery appears to slow with increasing $t_h$.  To quantify these trends, we fit the time-dependent strains using an empirical, stretched-exponential form, 
\begin{equation}
    \gamma(t_r) = 300\% - \big(\gamma_{r,0} +\Delta\gamma_r \left(1-\exp[-(t_r/\tau_r)^{\beta_r}]\right) \big),
    \label{stretched_exponential}
\end{equation}
where $\gamma_{r,0}$ is the magnitude of the initial, elastically recovered strain, $\Delta\gamma_r$ is the magnitude of the slow, glassy recovered strain at asymptotically late $t_r$, and $\tau_r$ and $\beta_r$ characterize the time scale and shape, respectively, of the glassy recovery.  The dashed line in Fig.~\ref{strainrec_3s} shows the result of a fit using Eq.~(\ref{stretched_exponential}), where the range of fitting is limited to $t_r>1$ s to exclude the damped oscillation at short $t_r$. Figure \ref{gamma_rec} shows the magnitudes $\gamma_{r,0}$ and $\Delta\gamma_r$ along with the total recoverable strain, $\gamma_r = \gamma_{r,0} + \Delta\gamma_r$, following different durations of stress relaxation.  The elastically recovered strain and the total recoverable strain decrease logarithmically with increasing hold time, as depicted by the solid line in Fig.~\ref{gamma_rec}, which show the result of a logarithmic fit to $\gamma_{r,0}$.  In contrast, the magnitude of the glassy strain recovery is essentially independent of hold time.  Thus, the stress relaxation following the flow cessation is accompanied by the conversion of recoverable strain to unrecoverable strain, where essentially all the loss of recoverable strain comes from the elastic component.  However, while the magnitude of recoverable strain associated with the glassy strain recovery is apparently insensitive to the preceding stress relaxation, the glassy recovery is not fully unaffected.   Most significantly, as shown in Fig.~\ref{beta_and_tau_rec}(a), the recovery time scale, $\tau_r$, increases strongly with increasing hold time.  In addition, as shown in Fig.~\ref{beta_and_tau_rec}(b), the shape of the glassy strain recovery, as parameterized by $\beta_r$, appears to evolve slightly with hold time.  However, we find that the strain recovery following different hold times can all be well fit using $\beta_r$ fixed to its average value, $\beta_r = 0.32$.  Results for $\gamma_{r}$, $\gamma_{r,0}$, $\Delta\gamma_r$, and $\tau_r$ obtained with $\beta_r$ fixed at 0.32 are also shown in Figs.~\ref{gamma_rec} and \ref{beta_and_tau_rec}(a).  This small value of $\beta_r$ reflects the strain recovery's highly stretched line shape. 

Since both the stress (Fig.~\ref{stressrelax}) and the recoverable strain (Fig.~\ref{gamma_rec}) decrease essentially logarithmically with hold time, the rates of change of the two are proportional,
\begin{equation}
    \frac{d\sigma}{dt_h} = G_{eff}\frac{d\gamma_{r}}{dt_h},
    \label{eq:sigmadot_v_gammadot}
\end{equation}
where the proportionality constant, $G_{eff} \approx 5800$ Pa, is an effective modulus.  For comparison, the linear storage modulus of the glass is $G' \approx 12000$ Pa, and the linear loss modulus is $G'' \approx 500$ Pa, as shown in Fig.~S2 the SM.  The value of $G_{eff}$ might reflect a transient reduction in the storage modulus following flow cessation due to strain softening, as predicted theoretically~\cite{MutnejaJoR2026}.  We also note that since the strain is held fixed during the stress relaxation, the decrease in the recoverable strain is accompanied by an equal and opposite increase in unrecoverable strain.  Thus, the rates of change of the stress and of the unrecoverable strain are related through a relationship equivalent to Eq.~(\ref{eq:sigmadot_v_gammadot}).  The fact that $G_{eff}$ falls between $G'$ and $G''$ points to contributions from both elastic and viscous stresses in driving the conversion of recoverable to unrecoverable strain.

\subsection{Microstructural Dynamics During Stress Relaxation}

\begin{figure}
\includegraphics[width=8cm]{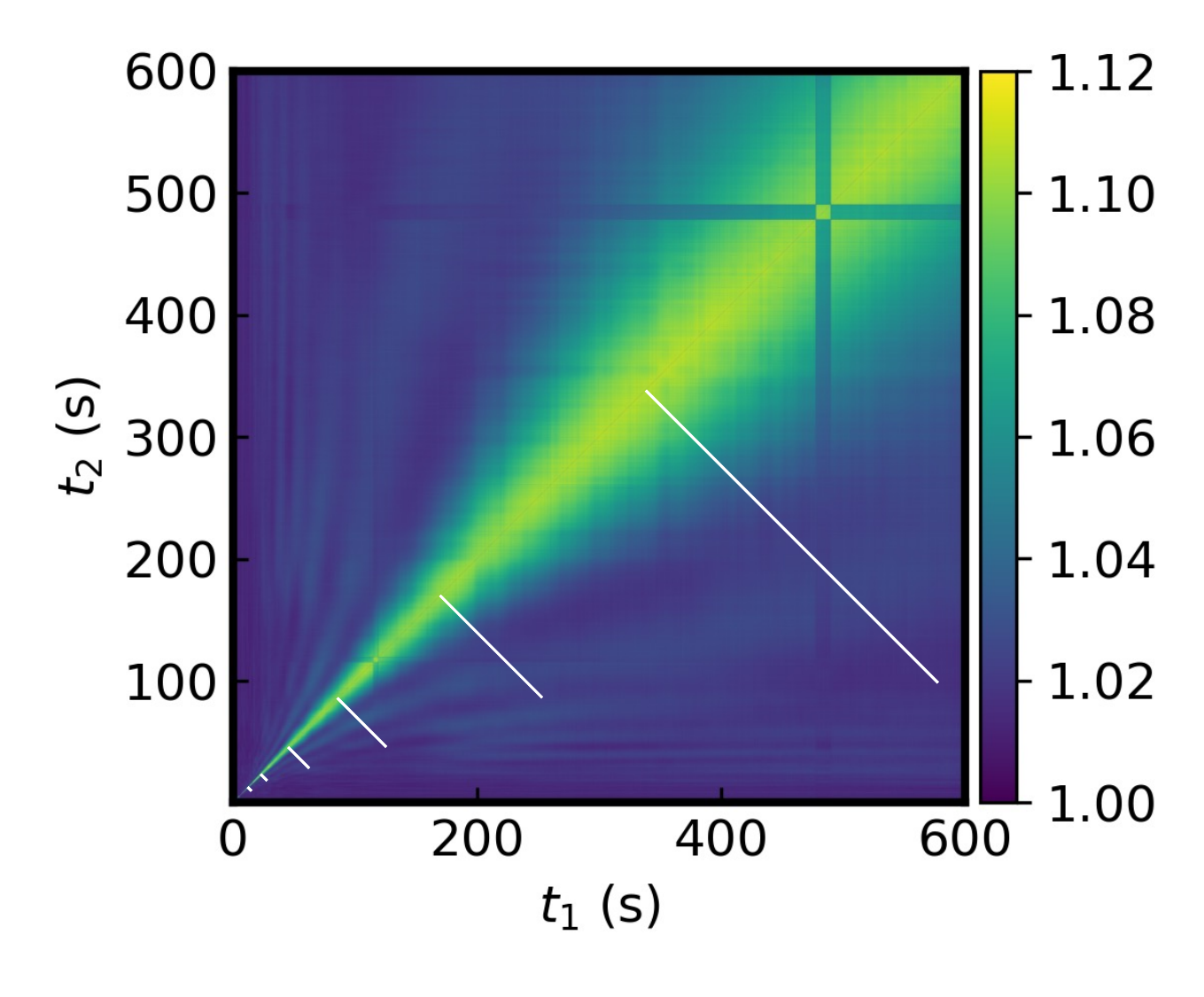}
\caption{Instantaneous correlation function $C(q,t_1,t_2)$ following cessation of shear flow to 300\% strain at 0.00316 s$^{-1}$ measured at $q = 0.25$ nm$^{-1}$ along the direction of the initial shear flow. The cessation of the flow ($t_h = 0$) is taken as the origin, $t_1 = t_2 = 0$.  The six white stripes indicate the regions included in calculating the intensity autocorrelation function $g_2(q,t| t_h)$ at $t_h =$ 11.5 s, 21.5 s, 42.5 s, 84.5 s, 168.5 s, and 336.5 s.}
\label{twotime_G0018}
\end{figure}

The microscopic dynamics during the stress relaxation captured by the XPCS measurements showed a strong dependence on hold time that is illustrated by the instantaneous correlation function~\cite{MadsenNJP2010}, 
\begin{equation}
C(\mathbf{q},t_1,t_2) = \frac{<I(\mathbf{q},t_1)I(\mathbf{q},t_2)>}{<I(\mathbf{q},t_1)><I(\mathbf{q},t_2)>},
\end{equation}
where $I(\mathbf{q},t)$ is the scattering intensity at wave vector $\mathbf{q}$ and time $t$, and the brackets indicate averages over detector pixels within a small vicinity of $\mathbf{q}$.  In the discussion below, we focus on $\mathbf{q}$ parallel to the direction of the initial shear flow.  Examples of results for $\mathbf{q}$ parallel to the vorticity direction are presented in the SM.  Figure \ref{twotime_G0018} shows the instantaneous correlation function at $q = 0.25$ nm$^{-1}$, which is near the first peak in the structure factor, during stress relaxation following shear flow to 300\% strain at  0.00316 s$^{-1}$.  The time when the strain reached 300\% is taken as the origin ($t_1 = t_2 = 0$) in Fig.~\ref{twotime_G0018}.  Shortly following the flow cessation, $C(q,t_1,t_2)$ is significantly larger than one only near the diagonal corresponding to small time differences ($t_1 \approx t_2$), indicating that the particle dynamics were initially rapid following the flow cessation.  With increasing hold time, the dynamics steadily slowed, and correlations persist for progressively larger time differences such that the band of large $C(q,t_1,t_2)$ values along the diagonal broadens.  

To analyze these dynamics quantitatively, we obtain the more familiar intensity autocorrelation functions $g_2(q,t|t_h)$ by averaging $C(q,t_1,t_2)$ at fixed delay time $t = |t_1-t_2|$,
\begin{equation}
g_2(q,t| t_h) =  \left< C(q,t_1,t_2)\right>_{t_h},
\end{equation}
where the average is over a small range of hold times so that the autocorrelation function can be considered an approximate snapshot of the evolving dynamics at a given $t_h$.  As an illustration of the procedure, Fig.~\ref{twotime_G0018} shows regions of the instantaneous correlation function indicated by white stripes that were included in calculating $g_2(q,t|t_h)$ at several $t_h$, as specified in the figure caption.  Each stripe is centered along a line of fixed $(t_1+t_2)/2$, which is taken as the hold time for that region, $t_h = (t_1+t_2)/2$.  To improve statistics, the width of each stripe (over which we average) spans one second, $t_h -0.5$ s $< t_h < t_h + 0.5$ s.  The length of each stripe is chosen to be as short as possible to minimize the effects of the evolving dynamics on the shape of $g_2(q,t| t_h)$ while still capturing the salient features of the correlation function.  

\begin{figure}
\includegraphics[width=6cm]{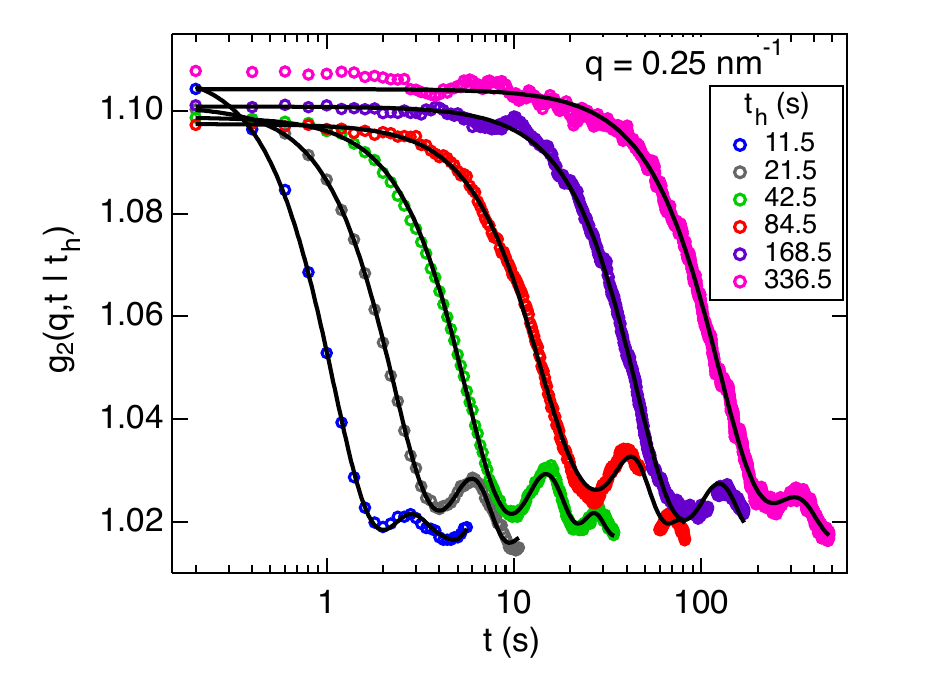}
\caption{Intensity autocorrelation functions at $q = 0.25$ nm$^{-1}$ along the flow direction at different hold times indicated in the legend following cessation of shear flow to 300\% strain at 0.00316 s$^{-1}$. The solid lines show results of fits based on the model described in the text (Eq.~(\ref{g2_hetero})).}
\label{G0018_g2s}
\end{figure}

\begin{figure}
\includegraphics[width=6cm]{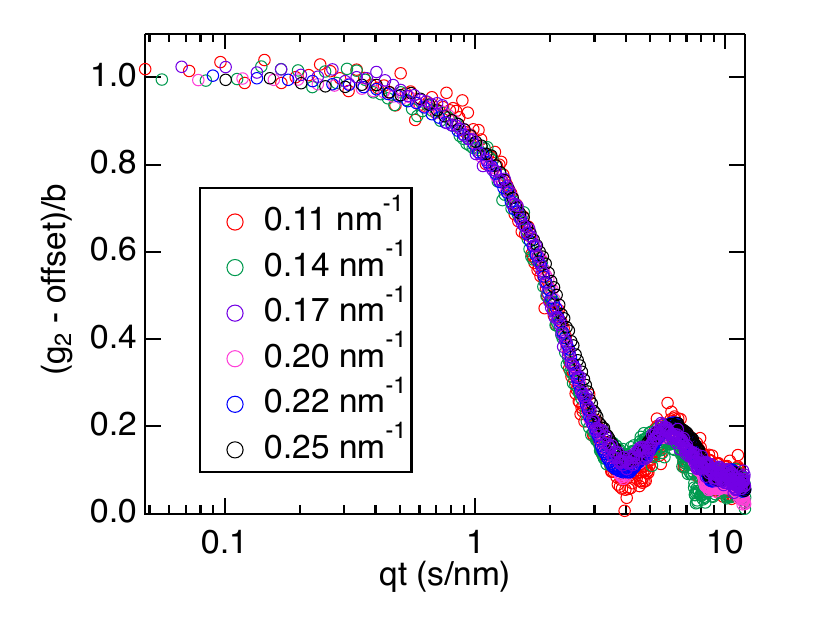}
\caption{Normalized intensity autocorrelation functions from the same measurement as in Fig.~\ref{G0018_g2s} at hold time $t_h = 60.5$ s and several wave vectors in the range 0.11 nm$^{-1}$ $< q < 0.25$ nm$^{-1}$ as indicated in the legend plotted against the product of delay time and wave vector.  The normalization involves subtracting a weakly $q$-dependent offset of approximately 1.01 from each correlation function and dividing by a weakly $q$-dependent contrast $b$.}
\label{G0018_g2scaled}
\end{figure}

Figure \ref{G0018_g2s} shows a set of autocorrelation functions obtained from  $C(q,t_1,t_2)$ in Fig~\ref{twotime_G0018}.  With increasing hold time, $g_2(q,t|t_h)$ decays at a later $t$, reflecting the steadily slowing microscopic dynamics.  Beyond the initial primary decay, the correlation function also shows a pronounced secondary peak whose size evolves with $t_h$.  This shape of $g_2(q,t|t_h)$ is strongly suggestive of heterodyning of the coherent scattering.  Heterodyning occurs when two regions within the coherently illuminated volume have distinct dynamics~\cite{LivetJSR2006}.  Several examples in the literature describe heterodyne signals in XPCS measurements and their analysis.  In some cases, the heterodyning results from coherent mixing of scattering from the sample of interest with that from a reference sample~\cite{LivetJSR2006,LhermitteRSI2017}.  In others, the dynamically distinct components reside within the same sample~\cite{UlbrandtNP2016,LewisRSI2018,MhannaJPCC2022,HorwathPRL2025}, which is the situation here.  That is, the XPCS line shapes in Fig.~\ref{G0018_g2s} indicate that during stress relaxation the glass is divided spatially into regions with different dynamical behavior.

To understand better the properties of the dynamics leading to the features in the correlation functions, in Fig.~\ref{G0018_g2scaled} we show $g_2(q,t|t_h)$ at $t_h = 60.5$ s and different $q$ normalized and plotted against delay time scaled by $q$.  The scaling leads to collapse of the correlation functions onto a single curve.  This scaling, which was observed for all initial shear rates and hold times, indicates that the different processes contributing to the decorrelation of $g_2(q,t|t_h)$ all involve dynamics in which relative particle displacements are linear in time. (As a counter example, one could imagine that diffusive processes dominate  $g_2(q,t|t_h)$, in which case the normalized correlation functions at different wave vector would collapse when plotted against $q^2t$.)

\begin{figure}
\includegraphics[width=6cm]{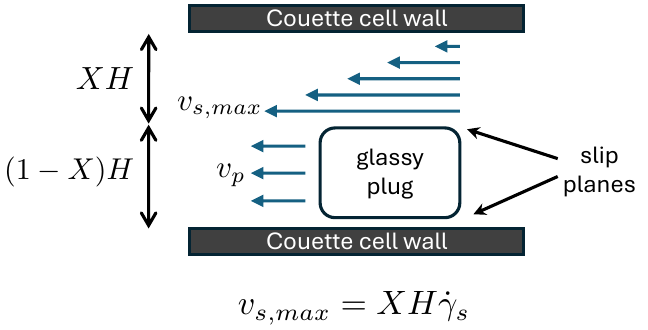}
\caption{Schematic depicting the glass's microscopic dynamics during stress relaxation following flow cessation. On spanning the Couette cell gap of size $H$, the glass is divided into two regions with distinct dynamics.  In one region that spans a fraction $X$ of the gap, the glass undergoes strain modeled as an affine deformation with maximum strain velocity $v_{s,max}$.  In the other, which spans $(1-X)$ of the gap, the glass moves as a plug with velocity $v_p$ while undergoing glassy relaxation.}
\label{schematic}
\end{figure}

Based on these two observations that (i) the shape of $g_2(q,t|t_h)$ reflects a heterodyne signal in which the scattering from parts of the glass with distinct dynamical properties mix coherently, and (ii) the different dynamical components share the property that the particle displacements are linear in time, we have constructed a model that accurately fits $g_2(q,t|t_h)$ at all $t_h$ following cessation of flow at the different shear rates investigated.  Figure \ref{schematic} shows a schematic depicting the model in which the dynamically distinct regions form two bands.  Such a separation of the sample into two dynamically distinct regions during stress relaxation could be a relic of shear-band formation during the preceding shear flow.  In one band, which comprises a fraction $X$ of the glass, the material undergoes strain-like motion that is qualitatively like the dynamics observed during stress relaxation following small step strains reported previously~\cite{chen2020microscopic}.  The dynamics in this band can be modeled with the intermediate scattering function for affine strain~\cite{Burghardt_PRE_2012}, 
\begin{equation}
g_{1,s}(q,t| t_h) =  \frac{\sin(q\dot{\gamma_s}XHt/2)}{q\dot{\gamma_s}XHt/2}.
\label{eq:g2affine}
\end{equation}
Here, $H = 200$ $\mu$m is the size of the Couette cell gap, and $\dot{\gamma_s}$ is the strain rate. Assuming a strain profile like that depicted schematically in Fig.~\ref{schematic}, the maximum strain velocity $v_{s,max}$ is related to the strain rate by $v_{s,max} = XH\dot{\gamma}_s$.  We also note that the wave vector in Eq.~(\ref{eq:g2affine}) technically refers to the component of the wave vector parallel to the strain.  Since we focus on the correlation functions at wave vectors parallel to the initial flow direction and the model assumes the strain during stress relaxation is anti-parallel to the initial flow, $q$ is simply the full wave-vector magnitude.

The other region of the glass, which comprises a fraction $(1-X)$, moves as a plug with velocity $\mathbf{v_p}$.  In addition, the glass undergoes slow, non-diffusive dynamics that are a characteristic of glassy materials and widely observed in XPCS and DLS measurements, and that are revealed by a compressed-exponential correlation function~\cite{CipellettiPRL2000,ChungPRL2006,RutaPRL2012,ConradPRE2015,DallariSciAdv2020}.  A model for these glassy dynamics, which is supported by simulation~\cite{BouzidNatPhys2017}, identifies them with the intermittent relaxation of random, internal stresses that become frozen into the glass upon solidification and predicts the commonly measured exponent for the compressed-exponential correlation function of 3/2~\cite{CipellettiFaraday2003}.   With both the plug flow and the glassy dynamics included, the second component can hence be modeled with the intermediate scattering function,  
\begin{equation}
g_{1,p}(q,t| t_h) =  \exp(-i\mathbf{q}\cdot\mathbf{v_p}t)\exp(-(t/\tau_g)^{3/2}),
\end{equation}
where $\tau_g$ characterizes the time scale of the glassy motion and scales inversely with wave vector, $\tau_g \sim q^{-1}$ \cite{CipellettiFaraday2003,ChungPRL2006,TrappePRE207,GuoPRE2007,ConradPRE2015, DallariSciAdv2020}.  Again, since we take the plug velocity to be anti-parallel to the initial flow velocity and we focus the analysis on wave vectors parallel to the initial flow, $\mathbf{q}\cdot\mathbf{v_p} = qv_p$.  The total intermediate scattering function is the weighted sum of the two components, $g_1(q,t| t_h) = Xg_{1,s}(q,t| t_h) + (1-X)g_{1,p}(q,t| t_h)$, and the resulting intensity autocorrelation function is obtained through the Siegert relation as~\cite{LivetJSR2006}
\begin{equation}
\begin{split}
g_2(q,t| t_h)  & = bg_1^2(q,t| t_h) + 1 \\
& = b \big( X^2g_{1,s}^2(q,t| t_h)\\
&+X(1-X)g_{1,s}(q,t| t_h)g_{1,p}(q,t| t_h)\\
&+X(1-X)g_{1,s}(q,t| t_h)g_{1,p}^*(q,t| t_h)\\
&+(1-X)^2g_{1,p}^2(q,t| t_h) \big) +1, 
\end{split}
\label{siegert}
\end{equation}
 since  $g_{1,s}(q,t| t_h)$ is real, leading to 
\begin{equation}
\begin{split}
g_2(q,t| t_h) = & b \Big( X^2\frac{\sin^2(q\dot{\gamma_s}XHt/2)}{(q\dot{\gamma_s}XHt/2)^2}\\
& +2X(1-X)\frac{\sin(q\dot{\gamma_s}XHt/2)}{q\dot{\gamma_s}XHt/2}\cos(qv_pt)\exp(-(t/\tau_g)^{3/2})\\
& + (1-X)^2\exp(-2(t/\tau_g)^{3/2}) \Big)+1,
\end{split}
\label{g2_hetero}
\end{equation}
where $b$ is the contrast.  The lines in Fig.~\ref{G0018_g2s} show the results of fits using Eq.~(\ref{g2_hetero}), where we also allow the baseline to have a small offset from 1.  Although this model is admittedly complicated, we find that it captures the main features of $g_2(q,t| t_h)$ at all $t_h$ following all initial shear rates and that removing any components from the model significantly degrades the quality of the fits.  We also note that $g_2(q,t| t_h)$ is insensitive to the positions of the affine strain and plug flow components within the gap; for instance, a flow profile in which the plug is in the center of the gap with affine strain at both walls or in which multiple plugs are separated by affine strain would also lead to Eq.~(\ref{g2_hetero}).  A discussion of how each component contributes to the total line shape and additional examples using Eq.~(\ref{g2_hetero}) to fit $g_2(q,t| t_h)$ following flow cessation at other shear rates are provided in the SM.

\begin{figure}
\includegraphics[width=6cm]{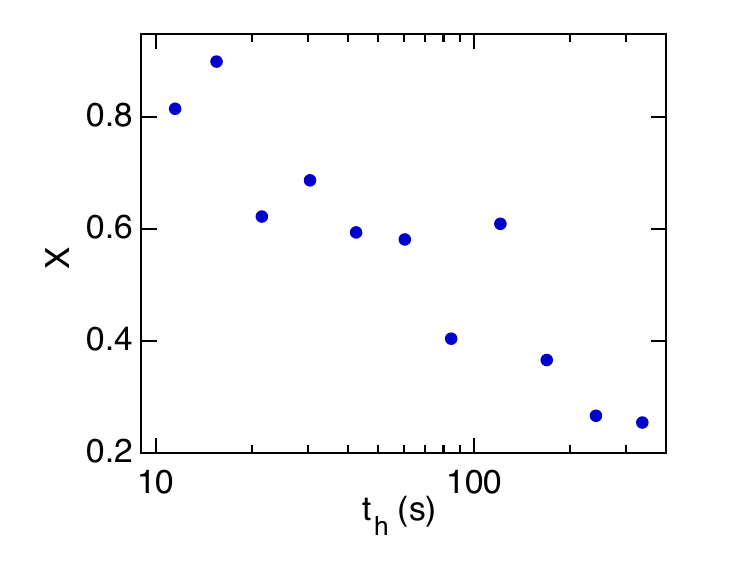}
\caption{Fraction of the glass that undergoes affine strain during stress relaxation as a function of hold time following cessation of shear flow to 300\% strain at 0.00316 s$^{-1}$. }
\label{X_vs_tw}
\end{figure}

\begin{figure}
\includegraphics[width=6cm]{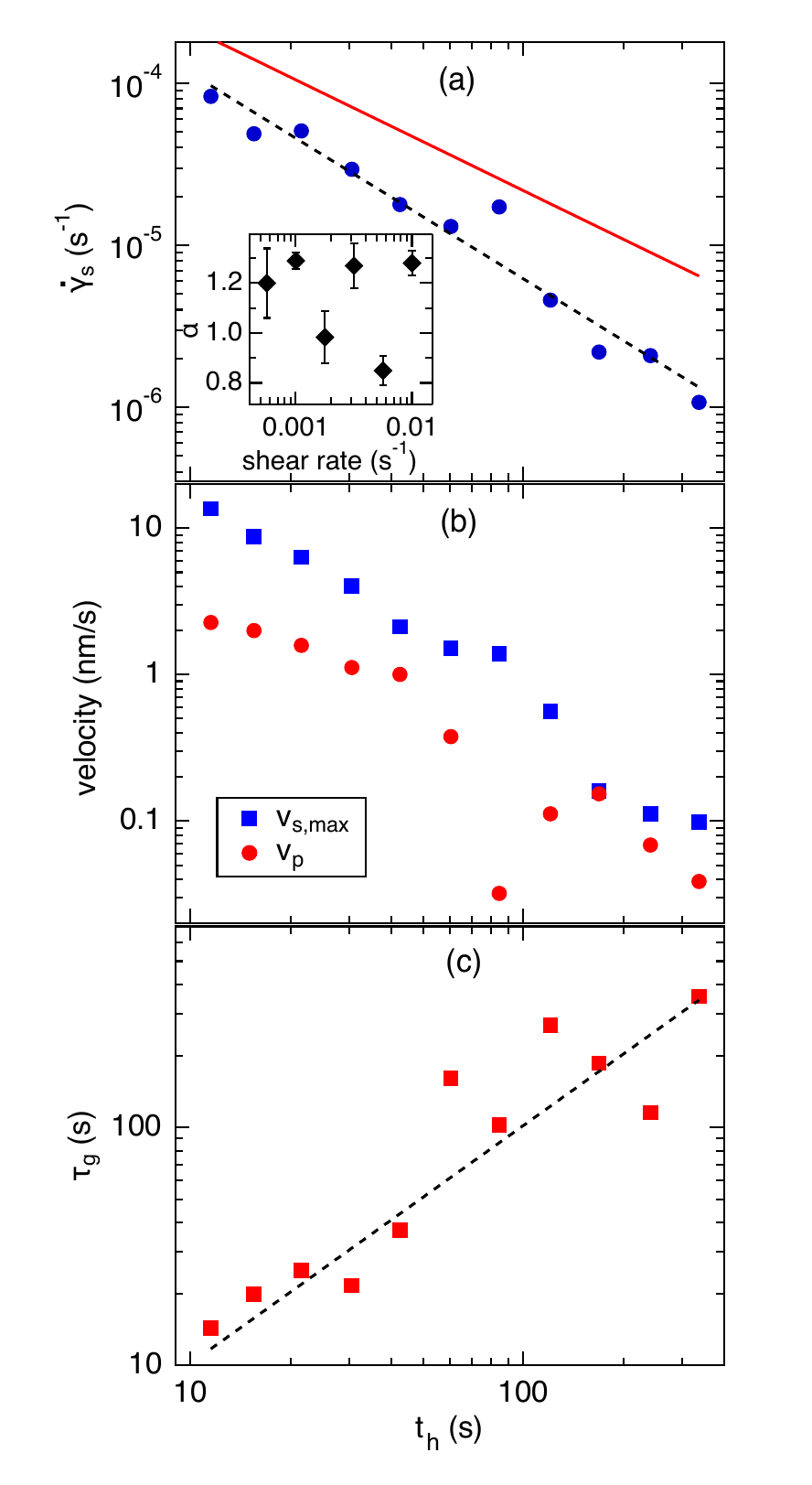}
\caption{(a) Strain rate of the affine deformation during stress relaxation as a function of hold time following cessation of shear flow to 300\% strain at 0.00316 s$^{-1}$.  The solid line depicts the rate of change of the recoverable strain obtained from the results in Fig.~\ref{gamma_rec} (see SM for details).  The dashed line shows the result of a power-law fit, $\dot{\gamma_s} \sim t_h^{-\alpha}$, and the inset shows the power-law exponent $\alpha$ at different initial shear rates. (b) Maximum velocity of the affine deformation (squares) and the plug flow velocity (circles) as functions of hold time from the same measurement.  (c) Correlation time (at $q=0.25$ nm$^{-1}$) of the glassy relaxation as a function of hold time from the same measurement.  The dashed line depicts a linear dependence, $\tau_g \sim t_h$.}
\label{gammadot_v_tau}
\end{figure}

Figure \ref{X_vs_tw} shows the fraction $X$ of the glass undergoing affine strain during the stress relaxation as a function of hold time following the cessation of flow at 0.00316 s$^{-1}$.  The fraction decreases with increasing hold time, a trend also observed following flow cessation at the other initial shear rates, as shown in Fig.~S6 in the SM.  Thus, as the stress relaxes, the profile of the convective backflow evolves so that plug-like motion comprises an increasing fraction of the glass.  Figure~\ref{gammadot_v_tau}(a) shows the rate $\dot{\gamma}_s$ of the affine strain component as a function of hold time.  Also shown in the figure by the solid line is an estimate for the rate of decrease of the recoverable strain based on the results in Fig.~\ref{gamma_rec}, as described in the SM.  Notably, the strain rate associated with the convective backflow measured with XPCS approximately tracks the rate of loss of recoverable strain measured with rheometry, indicating that the strain-like motion observed with XPCS is among the mechanisms contributing to the conversion of recoverable to unrecoverable strain during stress relaxation.  However, the observed strain rate is systematically lower than the rate at which recoverable strain is lost. Also, as shown in Fig.~\ref{gamma_rec}, the recoverable strain decreases with hold time approximately logarithmically, implying $\dot{\gamma}_r \sim t_h^{-1}$.  In contrast, the strain rate associated with the convective backflow shows a somewhat stronger power-law dependence on hold time.  The dashed line in Fig.~\ref{gammadot_v_tau}(a) shows the result of a power-law fit, $\dot{\gamma}_s \sim t_h^{-\alpha}$, which gives $\alpha = 1.27 \pm 0.09$.  The results following cessation of flow at the other initial shear rates show similar trends, as shown in Fig.~S8 of the SM.  As an illustration, the inset to Fig.~\ref{gammadot_v_tau}(a) shows the power-law exponent $\alpha$  at the different initial shear rates.  These differences between $\dot{\gamma}_s$ and $\dot{\gamma}_r$ point to additional mechanisms contributing to the loss of recoverable strain during stress relaxation beyond the affine strain, including contributions from the glassy relaxation and the plug flow that comprise the other components of the dynamics observed in the XPCS measurements.  

Figure \ref{gammadot_v_tau}(b) shows the plug velocity $v_p$ as a function of hold time following cessation of flow at 0.00316 s$^{-1}$.  For comparison, the maximum velocity of the strain component, $v_{s,max} = XH\dot{\gamma}_s$, is also plotted.  The plug velocity roughly tracks the maximum strain velocity but shows larger fluctuations with hold time, trends that again are shared by the results at other initial shear rates, as shown in Fig.~S9 in the SM.  As the figure in the SM shows, at lower initial shear rates $v_{s,max}$ and $v_p$ are similar in magnitude, while at higher initial shear rates, like 0.00316 s$^{-1}$ in Fig.~\ref{gammadot_v_tau}(b), $v_p < v_{s,max}$.  In either case, the fact that $v_p$ tracks $v_{s,max}$ implies that, like $\dot\gamma_s$, the plug velocity has a dependence on $t_h$ that approximately tracks the rate of decrease in recoverable strain, supporting the notion that the plug flow also contributes to the conversion of recoverable to unrecoverable strain during stress relaxation.  

Figure \ref{gammadot_v_tau}(c) shows the correlation time associated with the heterogeneous glassy relaxations from the same measurement. The dashed line in the plot depicts a linear dependence of $\tau_g$ on $t_h$.  Although the data are noisy, the time scale increases roughly linearly with hold time.  This hold-time dependence of $\tau_g$ is again common to the dynamics following cessation of flow at the other initial shear rates, as shown in Fig.~S7 in the SM.  Such linear growth in this correlation time has also been observed in other soft glassy materials following flow cessation~\cite{ChungPRL2006,BandyopadhyaySSC2006}.  This correlation time can be related to the inverse of the characteristic rates of random strain relaxations that comprise the glassy dynamics~\cite{CipellettiFaraday2003,BouzidNatPhys2017}.  Thus, the rate of this random motion decays approximately at $t_h^{-1}$, indicating these glassy dynamics also correlate with the rate of loss of recoverable strain.  As mentioned above, the loss of recoverable strain can more precisely be described as a conversion of recoverable strain to unrecoverable strain, with the two components having equal and opposite variations with hold time during the stress relaxation.  Thus, the correlations between the rate of loss of recoverable strain and the forms of microscopic dynamics underlying stress relaxation identified with XPCS can equivalently be considered as correlations between these dynamics and the rate of increase of unrecoverable strain.

\subsection{Microstructural Dynamics During Strain Recovery}

\begin{figure}
\includegraphics[width=6cm]{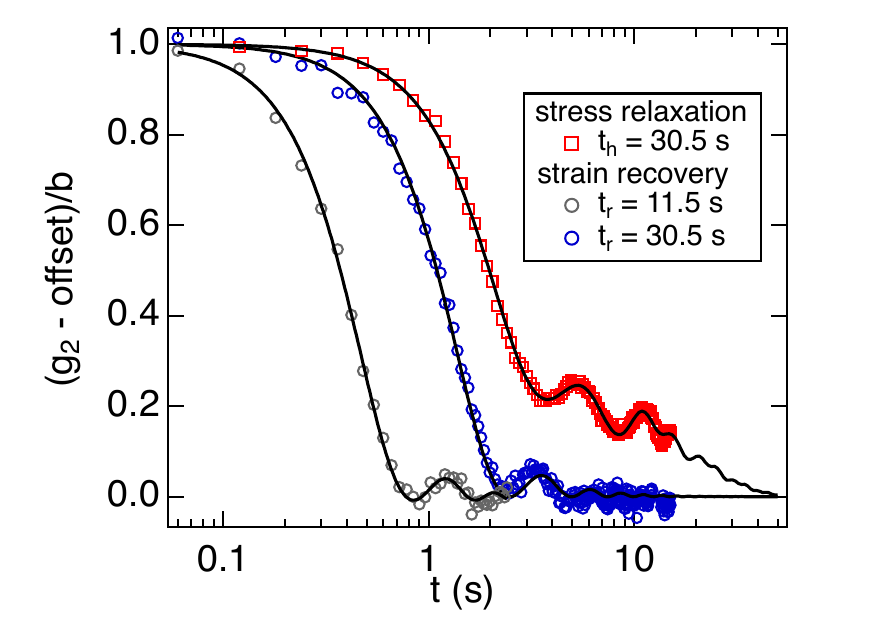}
\caption{Normalized intensity autocorrelation functions at $q = 0.25$ nm$^{-1}$ along the flow direction measured during strain recovery at recovery times $t_r = 11.5$ s (gray circles) and 30.5 s (blue circles) following cessation of shear flow to 300\% strain at 0.01 s$^{-1}$ and stress relaxation for 100 s. Also shown is the normalized intensity autocorrelation function measured during the stress relaxation at hold time $t_h = 30.5$ s (red squares).  The solid lines through the data during strain recovery show results of fits based on a model of simple affine shear strain (Eq.~(\ref{g2_affine})).  The solid line through the data taken during stress relaxation shows the result of a fit using the heterodyne model of spatially heterogeneous dynamics (Eq.~(\ref{g2_hetero})).}
\label{g2strainrecovery}
\end{figure}

\begin{figure}
\includegraphics[width=6cm]{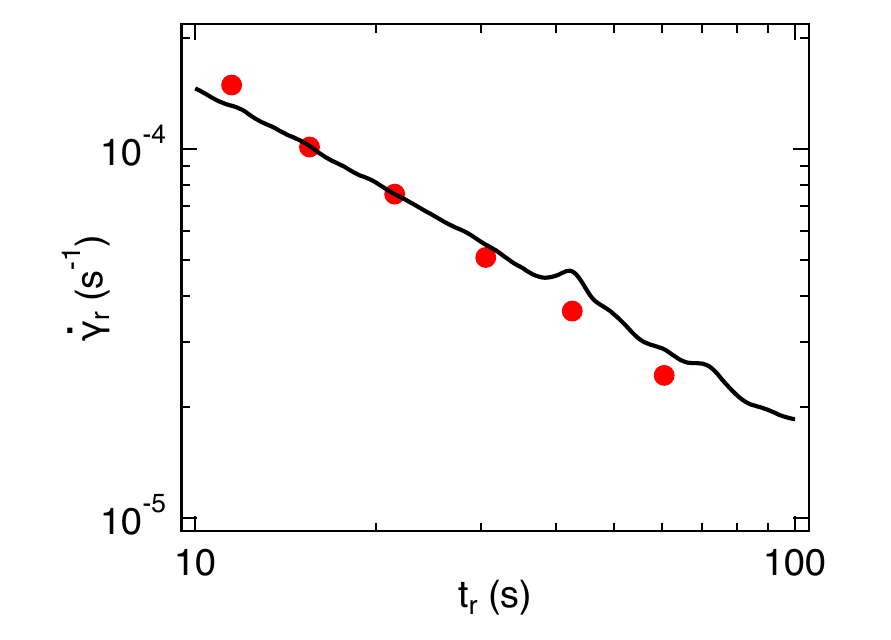}
\caption{Strain rates obtained from XPCS (red circles) during strain recovery as a function of recovery time following stress relaxation for $t_h = 100$ s.  The line shows the strain rate obtained directly from the rheometry measurement by numerically differentiating the recovered strain (Fig.~\ref{strainrec_alltw}). }
\label{strainrecovery_gammadot}
\end{figure}

The spatially heterogeneous microstructural picture required to model the dynamics during stress relaxation raises the question of whether similar dynamics occur during strain recovery.  To address this question, we examine the XPCS measurements obtained during strain recovery.  Figure \ref{g2strainrecovery} displays examples of the intensity autocorrelation function during strain recovery, $g_{2,r}(q,t| t_r)$, at two recovery times after the stress was set to zero following a hold time of $t_h = 100$ s along with an example intensity autocorrelation function measured during the preceding stress relaxation at $t_h = 30.5$ s.  Unlike the correlation functions during stress relaxtion, the correlation functions during strain recovery show no indication of a heterodyne signal that would imply spatially heterogeneous dynamics, and instead they are accurately modeled by a homogeneous affine strain. The lines through the data for $g_{2,r}(q,t| t_r)$ in Fig.~\ref{g2strainrecovery} show the results of fits using the lineshape for affine shear strain,
\begin{equation}
g_{2,r}(q,t| t_r) =  b\frac{\sin^2(q\dot{\gamma_r}Ht/2)}{(q\dot{\gamma_r}Ht/2)^2}+1
\label{g2_affine}
\end{equation}
Figure \ref{strainrecovery_gammadot} shows the strain rates $\dot{\gamma_r}$ obtained from such fits at different recovery times compared to the strain rate obtained by numerically differentiating the recovered strain as a function of $t_r$ from Fig.~\ref{strainrec_alltw}.  The results of the XPCS measurements match the rheometric strain rates well, further indicating that an affine strain profile accurately describes the dynamics during strain recovery.  To emphasize the difference in this behavior with that observed during the preceding stress relaxation, we also show in Fig.~\ref{g2strainrecovery} a fit to the correlation function during the stress relaxation using the line shape for heterogeneous dynamics (Eq.~(\ref{g2_hetero})).  The differences in line shapes during stress relaxation and during strain recovery make clear that the glass experiences distinct microscopic dynamics under the two conditions. 

\section{Conclusion}
\label{sec:conclusion}
This rheo-XPCS study on a nanocolloidal glass has provided several new insights into how yield stress fluids recover after flow cessation.  First, the correspondence between the rates of stress relaxation and loss of recoverable strain (Eq.~\ref{eq:sigmadot_v_gammadot}) points to the conversion of recoverable to unrecoverable strain as a mechanism by which rheological memory of the flow within the glass fades.  In addition, the similarities between the rate of decrease of recoverable strain and those of the components of the convective backflow indicate that the backflow is the primary microscopic process driving this conversion.  In light of this apparent significance of the convective backflow, an interesting question is why it takes on a spatially heterogeneous, banded structure.  One tempting speculation is that the heterogeneity becomes imprinted by shear banding during the preceding flow.  Indeed, simulations have reported correlations between flow characteristics and the motions associated with stress relaxation upon flow cessation~\cite{VinuthaPNAS2024}.  However, to test this idea experimentally, measurements that directly access the flow profile during the initial shear flow would be needed.  In principle, XPCS measurements could provide this information, but doing so would require using a different scattering geometry~\cite{ChenPoF2021} or smaller shear rates than those in the measurements presented here. In addition, rheo-XPCS experiments that compare the loss of recoverable strain with the stress relaxation and microscopic dynamics in other systems, such as colloidal gels~\cite{NegiPRE2009}, would be valuable.  

Two recent theoretical developments described above provide insight into the behavior observed here.  In the first, Lockwood and Fielding use the SGR model to construct a microscopic picture of the dynamics responsible for strain recovery following flow cessation that is similar to the strain recovery we observe in which a rapid, elastic contribution to the recoverable strain is followed by a slow, glassy contribution~\cite{LockwoodJoR2025}.  An interesting question is whether the SGR model can also provide insight into how recoverable strain is lost during stress relaxation at fixed strain as we described in Sec.~\ref{subsec:stressrelaxation} and as also reported in Ref.~\cite{NegiPRE2009}.  In the second, Mutneja and Schweizer~\cite{MutnejaJoR2026} have developed a theory employing a nonlinear Langevin equation to investigate stress relaxation in amorphous solids following step strains that captures the effects of convective backflow qualitatively like that we observed experimentally~\cite{chen2020microscopic}.  However, since both these theoretical approaches rely on mean-field-type approximations, spatial heterogeneity in the dynamics like we observe is seemingly outside the theories' scopes in their current forms.  An interesting question is whether extensions of these approaches could capture such spatially heterogeneous behavior.  Mutneja and Schweizer's theory further predicts that stress relaxation is accompanied by structural recovery that manifests in a relaxation of flow-induced distortions of the structure factor $S(q)$.  As shown in Fig.~S1(b) in the SM, we observe no measurable changes in the structure factor during the stress relaxation, but this absence of observed changes could reflect our restriction in the x-ray measurements to the structure in the flow-vorticity plane.  The relaxation of shear-induced distortions might, if present, more readily be seen in the flow-gradient plane~\cite{MohanJoR2015}.  The development of rheo-XPCS capabilities that can access this scattering plane would be valuable not only for addressing this question but also more broadly, since this geometry is considered the most informative in characterizing shear-induced structural changes in complex fluids in a range of contexts~\cite{CaputoMacromol2001,LiberatorePRE2006,EberleCOCIS2012}. Finally, we note several reports in the literature of non-monotonic stress relaxation in yield stress fluids following flow cessation~\cite{NegiJoR2010,HendricksPRL2019,SudreauPRM2022,WardPRM2025,KumarJoR2025,OwensPRL2025}, which in some cases has been ascribed to shear banding during the preceding flow or to aging occurring in parallel with the stress relaxation.  Rheo-XPCS measurements like those presented here to track the conversion of recoverable to unrecoverable strain and the microscopic dynamics underlying this intriguing behavior could be insightful.

\section{Acknowledgments}
We thank Mark Sutton for helpful discussions.  P.G.K. and S.A.R acknowledge support from the National Science Foundation under the DMREF Award Number DMR-25-22586.  J.L.H acknowledges support from the Natural Sciences and Engineering Research Council of Canada through the Discovery Grant award RGPIN-2024-06902.  The research used resources of the Advanced Photon Source and the Center for Nanoscale Materials, U.S. Department of Energy (DOE) Office of Science User Facilities operated for the DOE Office of Science by Argonne National Laboratory under Contract No. DE-AC02-06CH11357.

\appendix
   \section{Table of Symbols}
Table of symbols.

\begin{longtable}{l|l}
\caption{\label{tab:symbols}Table of symbols.}
\\ \hline 
   $\gamma$ &  {strain} \\
    \hline
    $\sigma$ &  {stress measured while the strain is held fixed at 300\% following shear flow} \\
    \hline
  $t_h$ &  {time that the strain has been held fixed at 300\% following shear flow} \\
   \hline
   $t_r$ &  {time since stress was set at zero in recovery rheology measurements} \\
  \hline
  $\gamma_{r,0}$ &  {magnitude of recoverable strain acquired elastically at $t_r \approx 0$} \\
  \hline
  $\Delta\gamma_r$ &  {magnitude of recoverable strain acquired during the slow, glassy strain recovery}\\
  \hline
  $\tau_r$ &  {time scale of the slow, glassy strain recovery} \\
  \hline
  $\beta_r$ &  {stretching exponent characterizing the slow, glassy strain recovery}\\
 \hline
 $\gamma_r$ &  {total recoverable strain}\\
 \hline
 $G_{eff}$ &  {effective modulus relating the rates of change of the stress and the recoverable strain}\\
 \hline
 $G'$ &  {linear elastic modulus}\\
 \hline
  $G''$ &  {linear loss modulus}\\
 \hline
 $\mathbf{q}$ &  {scattering wave vector}\\
 \hline
 $C(\mathbf{q},t_1,t_2)$ &  {XPCS instantaneous correlation function}\\
 \hline
 $g_1(q,t|t_h)$ &  {intermediate scattering function (ISF) during stress relaxation}\\
 \hline
  $g_{1,s}(q,t|t_h)$ &  {ISF for affine strain during stress relaxation}\\
 \hline
  $g_{1,p}(q,t|t_h)$ &  {ISF for plug flow and glassy relaxation during stress relaxation}\\
 \hline
 $g_2(q,t|t_h)$ &  {XPCS intensity autocorrelation function measured during stress relaxation}\\
 \hline
  $g_{2,r}(q,t|t_r)$ &  {XPCS intensity autocorrelation function measured during strain recovery}\\
 \hline
 $t$ &  {XPCS delay time}\\
 \hline
 $b$ &  {XPCS optical contrast}\\
 \hline
 $X$ &  {fraction of the glass undergoing strain motion during stress relaxation}\\
 \hline
 $1-X$ &  {fraction of the glass undergoing plug-like flow during stress relaxation}\\
 \hline
 $\dot{\gamma}_s$ &  {strain rate of the portion of the glass undergoing strain motion during stress relaxation}\\
 \hline
   $v_{s,max}$ &  {maximum strain velocity of the strain motion during stress relaxation}\\
   \hline
   $v_p$ &  {velocity of the portion of the glass undergoing plug flow during stress relaxation}\\
   \hline
   $\tau_g$ &  {relaxation time (at $q=0.25$ nm$^{-1}$) of the glassy dynamics during stress relaxation}\\
   \hline
   $\alpha$ &  {power-law exponent relating strain rate and hold time during stress relaxation}\\
     \hline
   $\dot{\gamma}_r$ &  {strain rate during strain recovery determined from XPCS}\\
   \hline\hline

\end{longtable}

\bibliography{bibliography}

\clearpage
\appendix

\renewcommand{\thefigure}{S\arabic{figure}}
\setcounter{figure}{0}

\section*{Supplemental Material}
\section{Structure of the nanocolloidal glass}

\begin{figure}[!t]
    \centering
    \includegraphics[width = 2.9in]{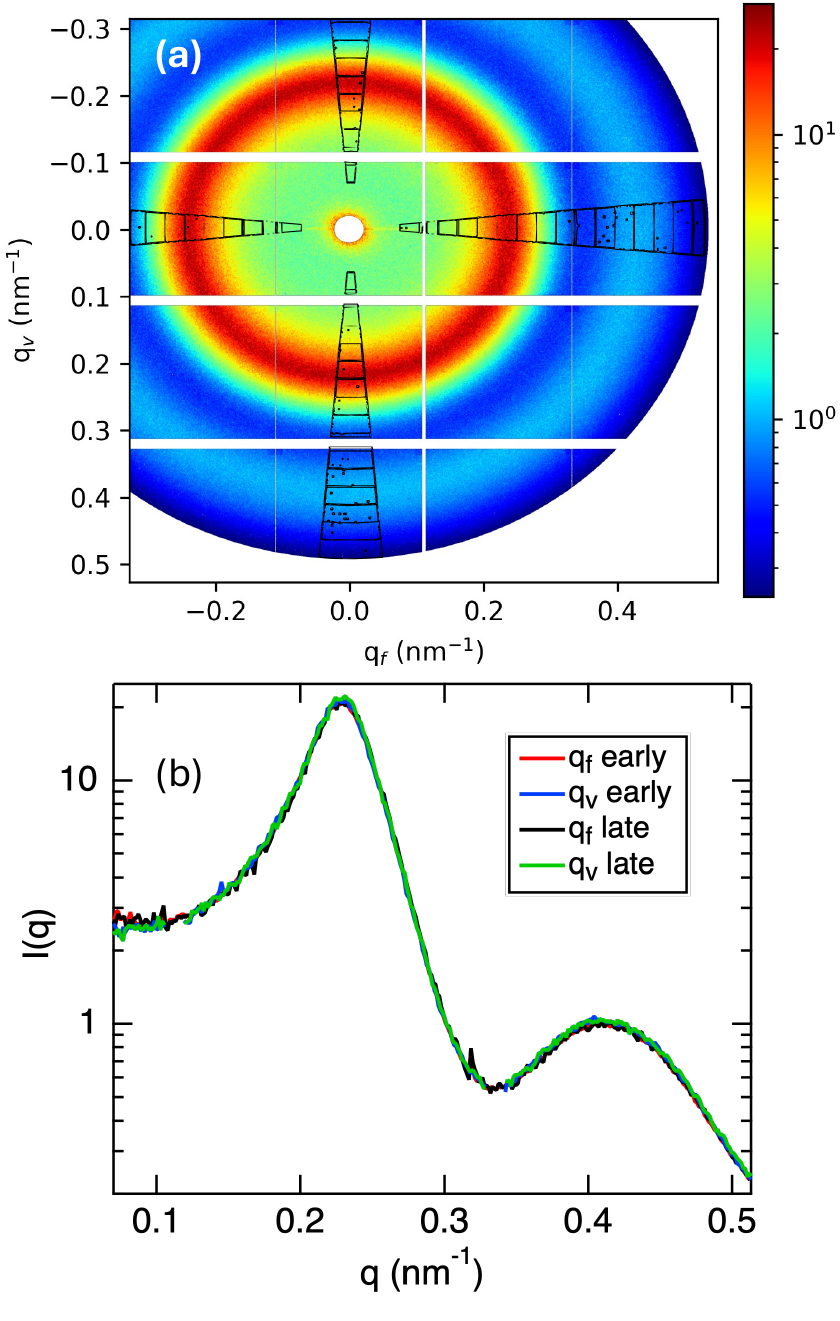}
    \caption{(a) Average of 10$^4$ x-ray scattering images obtained at 10 frames per second following  cessation of shear flow to 300\% strain at 0.00316 s$^{-1}$. The scattering intensity is expressed as the number of photons per detector pixel as a function of wave vector along the flow ($q_f$) and vorticity ($q_v$) directions. The regions outlined in black indicate the pixels over which the time correlations were averaged to obtain $C(\mathbf{q},t_1,t_2)$ at 16 different $|\mathbf{q}|$ along the flow and vorticity directions.  (b) Scattering intensity $I(q)$ along the flow and vorticity directions obtained from averaging over 100 seconds immediately following the cessation of flow (early) and 15 minutes after the cessation (late).}   
    \label{scattering}
\end{figure}

Figure \ref{scattering}(a) shows the average of 10$^4$ x-ray scattering images obtained at 10 frames per second following the cessation of shear flow to 300\% strain at 0.00316 s$^{-1}$.   Also shown on the image are partitions denoting the pixels over which the time correlations were averaged to obtain $C(\mathbf{q},t_1,t_2)$ at sixteen different $|\mathbf{q}|$ parallel to the initial flow direction ($q_f$) and parallel to the vorticity direction ($q_v$). Figure \ref{scattering}(b) shows the scattering intensity $I(q)$ averaged over 1000 images during the first 100 seconds following cessation of shear flow to 300\% strain at 0.00316 s$^{-1}$ and averaged over 1000 images approximately 15 minutes after the cessation of shear flow.  Results are shown separately for wave vectors along the flow and vorticity directions.  $I(q)$ has a large peak near $q$ = 0.23 nm$^{-1}$ corresponding to a structure-factor peak associated with near neighbor spatial correlations in the glass. The identical shapes of $I(q)$ along the different wave vector directions and at different times indicate that (in the flow-vorticity plane) no anisotropies in the average structure developed during the shear flow and no measurable changes to the average structure occurred during the stress relaxation.

\section{Rheology of the glass}
\begin{figure}
\centering
\includegraphics[width = 2.9in]{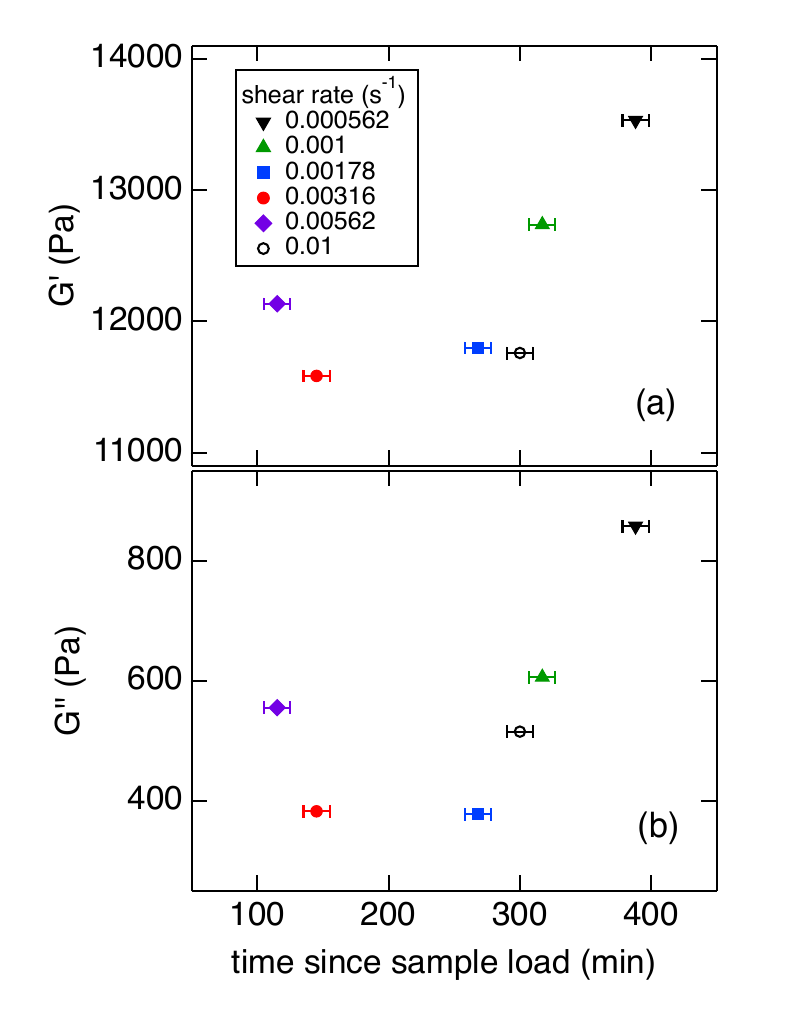}
\caption{(a) Storage and (b) loss modulus at 1 rad/s measured prior to each flow cessation experiment as a function of the time since the sample was loaded in the rheometer.  The legend indicates the shear rate of each flow cessation experiment. Except for the experiment with shear rate 0.01 s$^{-1}$, all of the experiments were performed on the same load of sample in the rheometer.}
\label{shearmoduli}
\end{figure}

As described in the main text, prior to each flow cessation experiment, the glass was subjected to a preshear protocol followed by a measurement of the linear shear modulus.  Figure \ref{shearmoduli} displays the linear storage modulus $G'$ and loss modulus $G''$ at 1 rad/s measured prior to each experiment as a function of the time since the sample was loaded in the rheometer.  All flow cessation experiments with the exception of the one with shear rate of 0.01 s$^{-1}$  were performed using the same load of sample in the rheometer.  As Fig.~\ref{shearmoduli} depicts, the moduli were stable until about 300 minutes after loading, when they started to increase slightly, perhaps due to a small amount of evaporation of water from the colloidal suspension.  As a precaution, a fresh sample was therefore loaded in the rheometer, and the experiment with shear rate of 0.01 s$^{-1}$ was performed.  All the recovery rheology experiments were also performed with this second load of sample within 300 minutes of the loading.  Notably, the moduli showed good reproducibility between sample loads.

\section{Heterodyne lineshape}
\begin{figure}
    \centering
    \includegraphics[width = 2.9in]{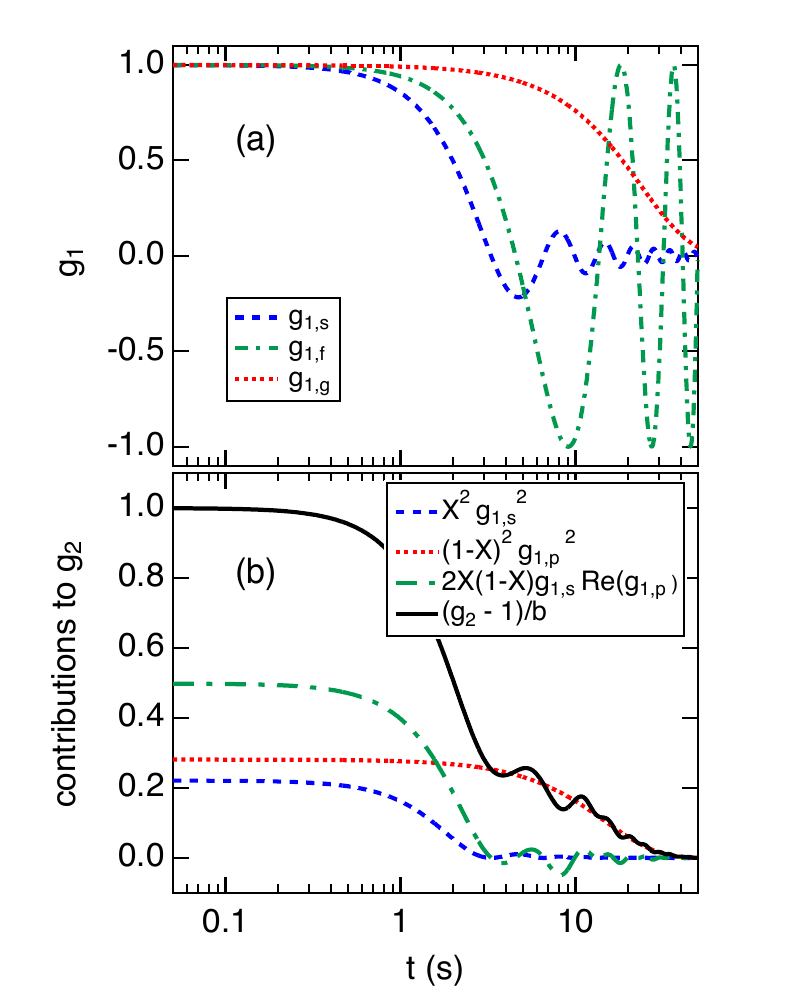}
    \caption{(a) Intermediate scattering functions corresponding to the affine strain (blue dashed line), plug flow (green dash-dotted line), and glassy (red dotted line) contributions to the dynamics during stress relaxation.  The functions are based on Eqs.~(\ref{eq:g1affine})-(\ref{eq:g1cipelletti}) with the following parameters: $\dot{\gamma}_s = 8.1\times10^{-5}$ s$^{-1}$, $v_p = 1.4$ nm/s, and $\tau_g =$ 23.6 s.  (b) Normalized heterodyne intensity autocorrelation function and the three terms comprising it in Eq.~(\ref{g2_hetero}), as specified in the legend.  The parameters are the same as in (a) along with $X =$ 0.47.}
    \label{heterodyne}
\end{figure}

As described in the main text, the XPCS intensity autorcorrelation function during stress relaxation reflects a heterodyne scattering signal with three contributions to the dynamics -- an affine strain, a plug flow, and a glassy relaxation -- that combine to give Eq.~(8).  Here, we provide an illustrative example of how each contributes to the correlation function.   When $\mathbf{q}$ is along the flow direction, the intermediate scattering function for each term is given by, 
\begin{equation}
g_{1,s}(q,t| t_h) =  \frac{\sin(q\dot{\gamma_s}XHt/2)}{q\dot{\gamma_s}XHt/2}.
\label{eq:g1affine}
\end{equation}
\begin{equation}
g_{1,f}(q,t| t_h) =  \exp(-iqv_pt),
\label{eq:g1plug}
\end{equation}
\begin{equation}
g_{1,g}(q,t| t_h) = \exp(-(t/\tau_g)^{3/2}),
\label{eq:g1cipelletti}
\end{equation}
where the subscripts $s$, $f$, and $g$ refer to affine strain, plug flow, and glassy relaxation, respectively, and $q$ is assumed to be parallel (or anti-parallel) to the strain and plug flow velocities.  Each of these terms is shown in Fig.~\ref{heterodyne}(a) at $q =$ 0.25 nm$^{-1}$, where the real part of $g_{1,f}(q,t| t_h)$ is plotted, and the parameter values are chosen to match those from the fit to $g_2(q,t| t_h)$ at $t_h = 30.5$ s shown in Fig.~13 of the main text:   $\dot{\gamma}_s = 8.1\times10^{-5}$ s$^{-1}$, $v_p = 1.4$ nm/s, and $\tau_g =$ 23.6 s.  As described in the main text, the heterodyne signal comes from the combination of coherent scattering from two regions of the glass, one of which undergoes affine strain and the other of which undergoes plug flow. For simplicity, the glassy relaxation is assumed to contribute only in the region undergoing plug flow, which hence leads to an intermediate scattering function for that region of,
\begin{equation}
\begin{split}
g_{1,p}(q,t| t_h) & = g_{1,f}(q,t| t_h)g_{1,g}(q,t| t_h)\\
& = \exp(-iqv_pt-(t/\tau_g)^{3/2})
\end{split}
\label{eq:g1glassyplug}
\end{equation}
The total intermediate scattering function is then the weighted sum of the coherent scattering from the two regions,
\begin{equation}
g_{1}(q,t| t_h) = Xg_{1,s}(q,t| t_h) + (1-X)g_{1,p}(q,t| t_h), 
\end{equation}
where $X$ is the fraction of the glass undergoing affine strain.  The resulting intensity autocorrelation function is obtained through the Siegert relation as~\cite{LivetJSR2006}
\begin{equation}
\begin{split}
g_2(q,t| t_h)  & = bg_1^2(q,t| t_h) + 1 \\
& = b \big( X^2g_{1,s}^2(q,t| t_h)\\
&+X(1-X)g_{1,s}(q,t| t_h)g_{1,p}(q,t| t_h)\\
&+X(1-X)g_{1,s}(q,t| t_h)g_{1,p}^*(q,t| t_h)\\
&+(1-X)^2g_{1,p}^2(q,t| t_h) \big) +1.
\end{split}
\label{eq:siegert1}
\end{equation}
since $g_{1,s}(q,t| t_h)$ is real, or
\begin{equation}
\begin{split}
&g_2(q,t| t_h) =  b \Big( X^2\frac{\sin^2(q\dot{\gamma_s}XHt/2)}{(q\dot{\gamma_s}XHt/2)^2}\\
& +2X(1-X)\frac{\sin(q\dot{\gamma_s}XHt/2)}{q\dot{\gamma_s}XHt/2}\cos(qv_pt)\exp(-(t/\tau_g)^{3/2})\\
& + (1-X)^2\exp(-2(t/\tau_g)^{3/2}) \Big)+1,
\end{split}
\label{g2_hetero}
\end{equation}
where $b$ is the contrast.  The three terms on the right side of Eq.~(\ref{g2_hetero}) along with their sum, $(g_2(q,t| t_h) - 1)/b$, are plotted in Fig.~\ref{heterodyne}(b), where again the parameter values are chosen to match those from the fit to $g_{2}(q,t| t_h)$ at $t_h = 30.5$ s shown in Fig.~13 of the main text, where $X = 0.47$.

\section{Dynamics in vorticity direction}
\begin{figure}
\includegraphics[width=2.9in]{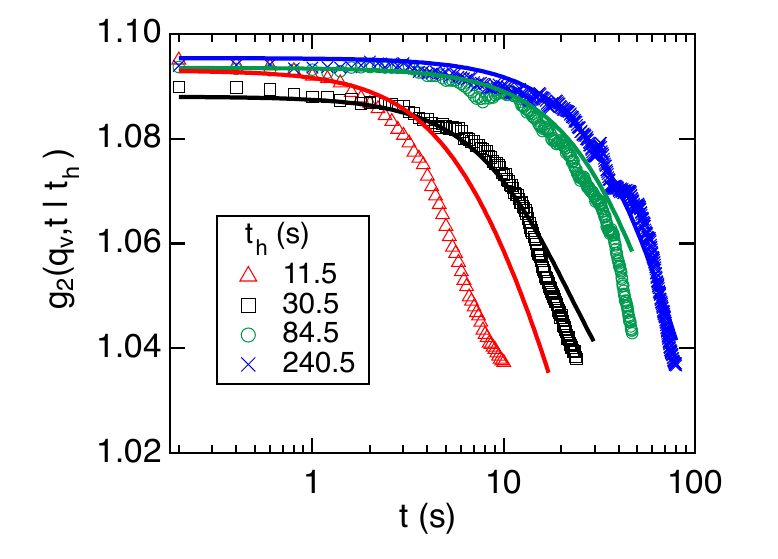}
\caption{Intensity autocorrelation functions from the same measurement that produced the data in Fig.~8 of the main text at $ q = 0.25$ nm$^{-1}$ along the vorticity direction and at hold times indicated in the legend.  The solid lines show predictions for the correlation functions based on fits to the correlation functions with $\mathbf{q}$ along the flow direction as described in the text.}
\label{g2_vorticity}
\end{figure}

Figure \ref{g2_vorticity} shows a set of correlation functions with $\bf{q}$ along the vorticity direction, $q_v$, at $q_v = 0.25$ nm$^{-1}$ from the same measurement that produced the data in Fig.~8 of the main text. As a comparison with Fig.~8 shows, the correlation functions with $\bf{q}$ along the vorticity direction decay on a longer time scale than those with $\bf{q}$ along the flow direction. This difference is expected based on the model to fit the correlations functions with $\bf{q}$ along the flow direction. The wave vectors that enter the expressions for the intermediate scattering function for the affine strain and plug flow, Eqs.~(\ref{eq:g1affine}) and (\ref{eq:g1plug}), refer the components of the wave vectors along the affine strain and plug flow motion, which is taken to be anti-parallel to the initial flow direction.  Hence, these terms contribute to the decay of $g_2(q,t|t_h)$ only to the extent that the wave vector has a component in that direction.  To get a sense of whether the results with $\bf{q}$ along the vorticity direction are consistent with the model, we show with the solid lines in Fig.~\ref{g2_vorticity} predictions for the correlation functions in the vorticity direction based on the analysis presented in the main text for the results with $\bf{q}$ along the flow direction.  To calculate these predictions, we use the fact that the region of the area detector over which the average is taken to determine $g_2(q,t|t_h)$ subtends an angle $\pm 5^\circ$ to approximate the angle between $\bf{q}$ and the strain and plug motions when $\bf{q}$ is nominally along the vorticity direction to be $87.5^\circ$.  (See Fig.~\ref{scattering}(a).) That is, to calculate the lines in Fig.~\ref{g2_vorticity}, we use Eq.~\ref{g2_hetero} with $q = |\mathbf{q}|\cos(87.5^\circ)  \approx 0.011$ nm$^{-1}$ and the parameter values ($b$, $X$, $\dot{\gamma}_s$, $v_p$, and $\tau_g$) from the fits to $g_2(q,t|t_h)$ along the flow direction shown in Fig.~8 of the main text.  As Fig.~\ref{g2_vorticity} shows, the calculated correlation functions provide a good approximation for the measured correlation functions along the vorticity direction, providing further support for the model describing the dynamics.  Notably, this success indicates no other processes contribute significantly to the dynamics along the vorticity direction.  This finding contrasts with our previous study of the dynamics associated with stress relaxation following smaller step strains, where intermittent, avalanche-like events dominated the motion in the vorticity direction~\cite{chen2020microscopic}.  

\section{Correlation functions following cessation of flow at different shear rates}

\begin{figure*}
\includegraphics[width=16cm]{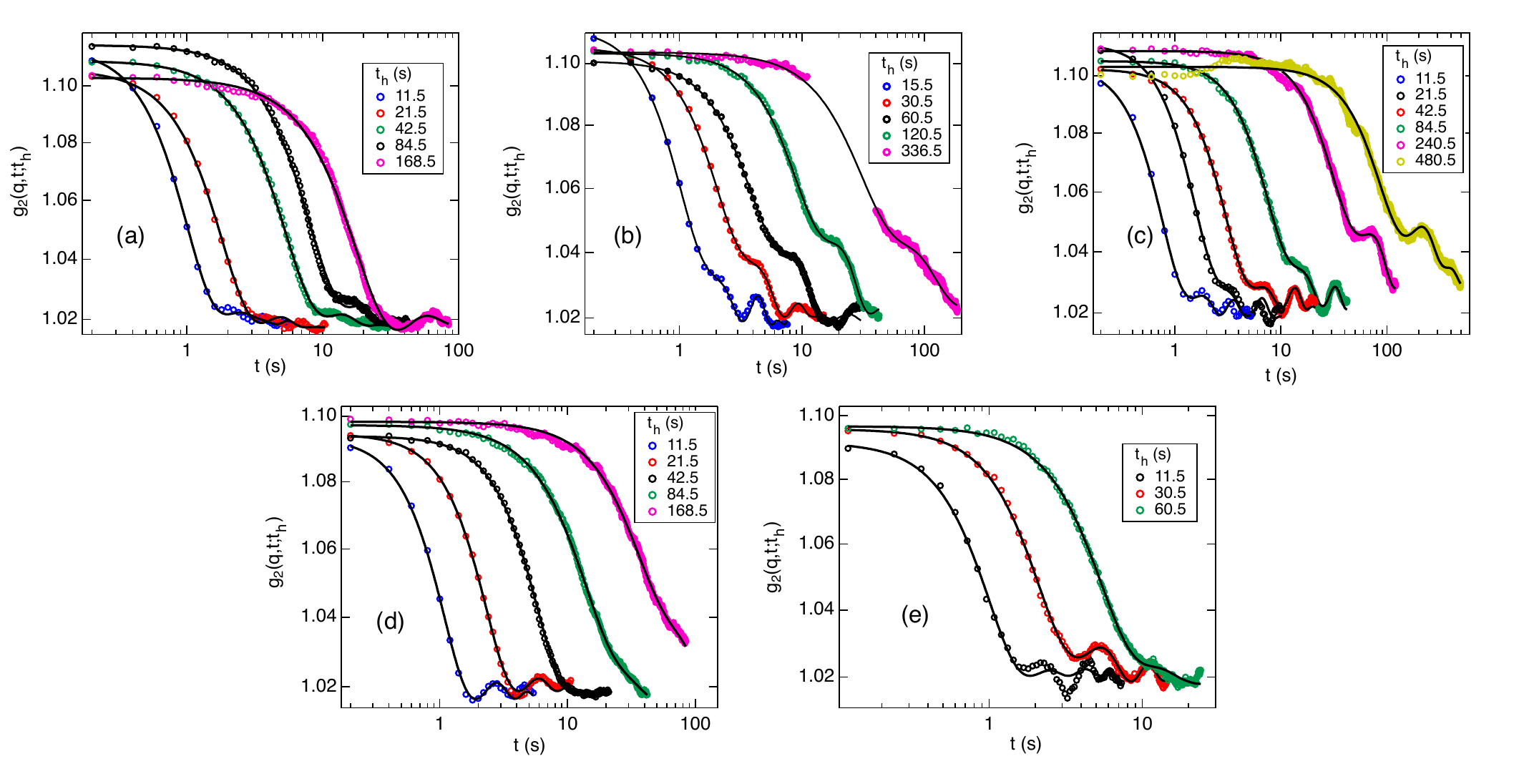}
\caption{Intensity autocorrelation functions at $q = 0.25$ nm$^{-1}$ along the flow direction at different at hold times indicated in the legends following cessation of shear flow to 300\% strain at shear rates of (a) 0.000562 s$^{-1}$, (b) 0.001 s$^{-1}$, (c) 0.00178 s$^{-1}$, (d) 0.00562 s$^{-1}$, and (e) 0.01 s$^{-1}$.  The solid lines show the results of fits using the model described in the main text (Eq.~(8)).}
\label{g2s}
\end{figure*}

As shown in Fig.~1 of the main text, measurements were performed following cessation of shear flow to 300\% strain at six shear rates:  0.000562, 0.001, 0.00178, 0.00316, 0.00562, and 0.01 s$^{-1}$.  In the main text, an analysis of the XPCS results from the measurement following cessation of flow at 0.00316 s$^{-1}$ was presented as a representative example.  Here we show the results of analysis of the intensity autocorrelation functions obtained following flow cessation at the other shear rates.  
Figure \ref{g2s} shows a set of correlation functions at different hold times following shear flow at the different initial shear rates.  The solid lines in the figure show results of fits using the model described above and in the main text.  

\begin{figure*}[t]
\centering
\begin{minipage}[t]{\columnwidth}
\centering
\includegraphics[width=2.9in]{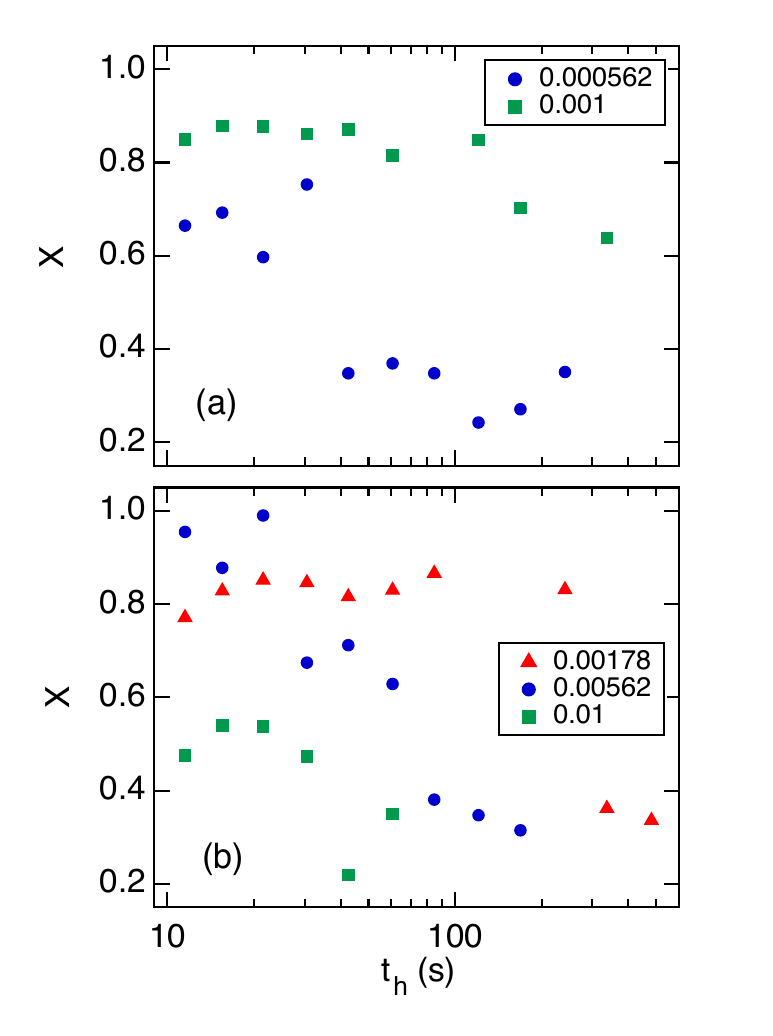}
\caption{Fraction of the glass that undergoes affine strain during stress relaxation as a function of hold time following cessation of shear flow to 300\% strain at (a) 0.000562 \&  0.001 s$^{-1}$, and at (b) 0.00178, 0.00562 \& 0.01 s$^{-1}$.}
\label{X}
\end{minipage}
\hfill
\begin{minipage}[t]{\columnwidth}
\centering
\includegraphics[width=2.9in]{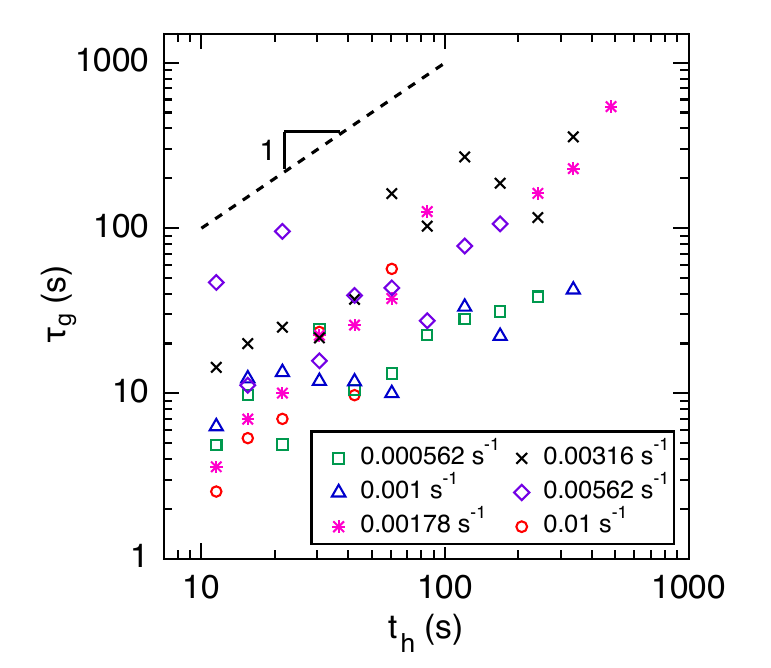}
\caption{Time scale of the glassy relaxation (at $q = 0.25$ nm$^{-1}$) as a function of hold time following cessation of flow to 300\% strain at shear rates indicated in the legend.  The dashed line depicts a linear dependence, $\tau_g \sim t_h$.}
\label{tau}
\end{minipage}
\end{figure*}

Figure \ref{X} shows the fraction $X$ of the glass undergoing affine strain during the stress relaxation as a function of hold time following cessation of flow at the different rates, and Fig.~\ref{tau} shows the characteristic time of the glassy relaxation as a function of hold time following cessation of flow at the different rates.  The dashed line in Fig.~\ref{tau} depicts a linear dependence on hold time, $\tau_g \sim t_h$.  

\begin{figure*}[t]
\includegraphics[width=16cm]{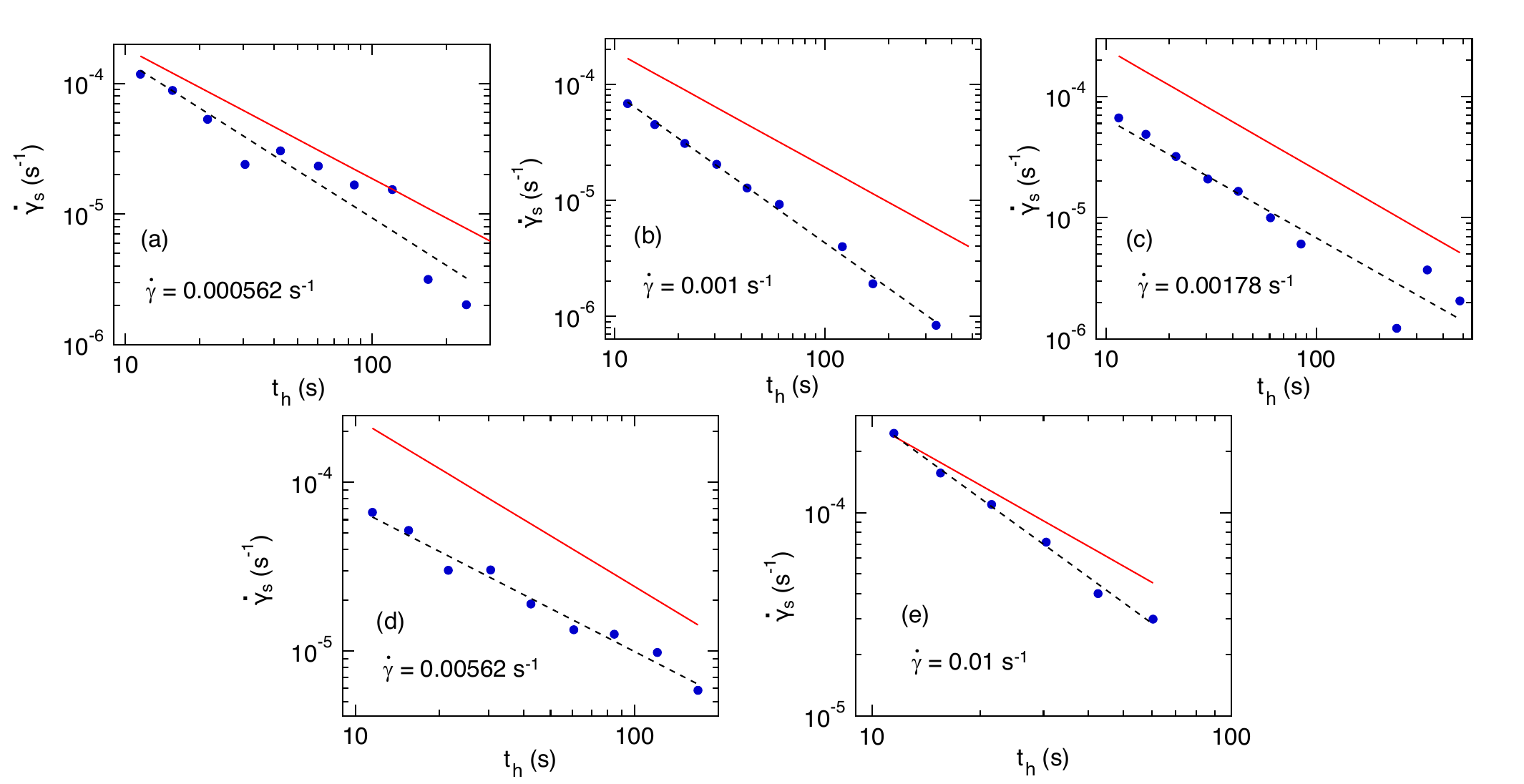}
\caption{Strain rate of the affine deformation during stress relaxation as a function of hold time following cessation of shear flow to 300\% strain at shear rates of (a) 0.000562 s$^{-1}$, (b) 0.001 s$^{-1}$, (c) 0.00178 s$^{-1}$, (d) 0.00562 s$^{-1}$, and (e) 0.01 s$^{-1}$.  The solid lines depict the rate of change of the recoverable strain, as described in the text.  The dashed lines show the results of power-law fits, $\dot{\gamma_s} \sim t_h^{-\alpha}$.  The power-law exponents $\alpha$ are shown in the inset to Fig.~12(a).}
\label{gammadots}
\end{figure*}

Figure \ref{gammadots} shows the strain rate of the affine strain component as a function of hold time following cessation of flow at the different rates.  The dashed line in each panel shows the result of a power-law fit to the strain rate, and the solid line depicts an estimate of the rate of loss of recoverable strain.  Since the recoverable strain as a function of hold time was measured only following initial shear flow at 0.01 s$^{-1}$ (Fig.~5 of the main text), to obtain the estimates of the rate of loss of the recoverable strain following flows at the other initial shear rates, we used the rate of decrease of the stress following each shear rate (Fig.~2(b) of the main text) along with the relation between between the rate of change of the stress and the rate of change of the recoverable strain (Eq.~(2) of the main text), where we assumed the proportionality constant, $G_{eff} = 5800$ Pa, was the same for all initial shear rates.  This process was also employed to obtain the solid line in Fig.~12(a) of the main text.

\begin{figure*}
\includegraphics[width=16cm]{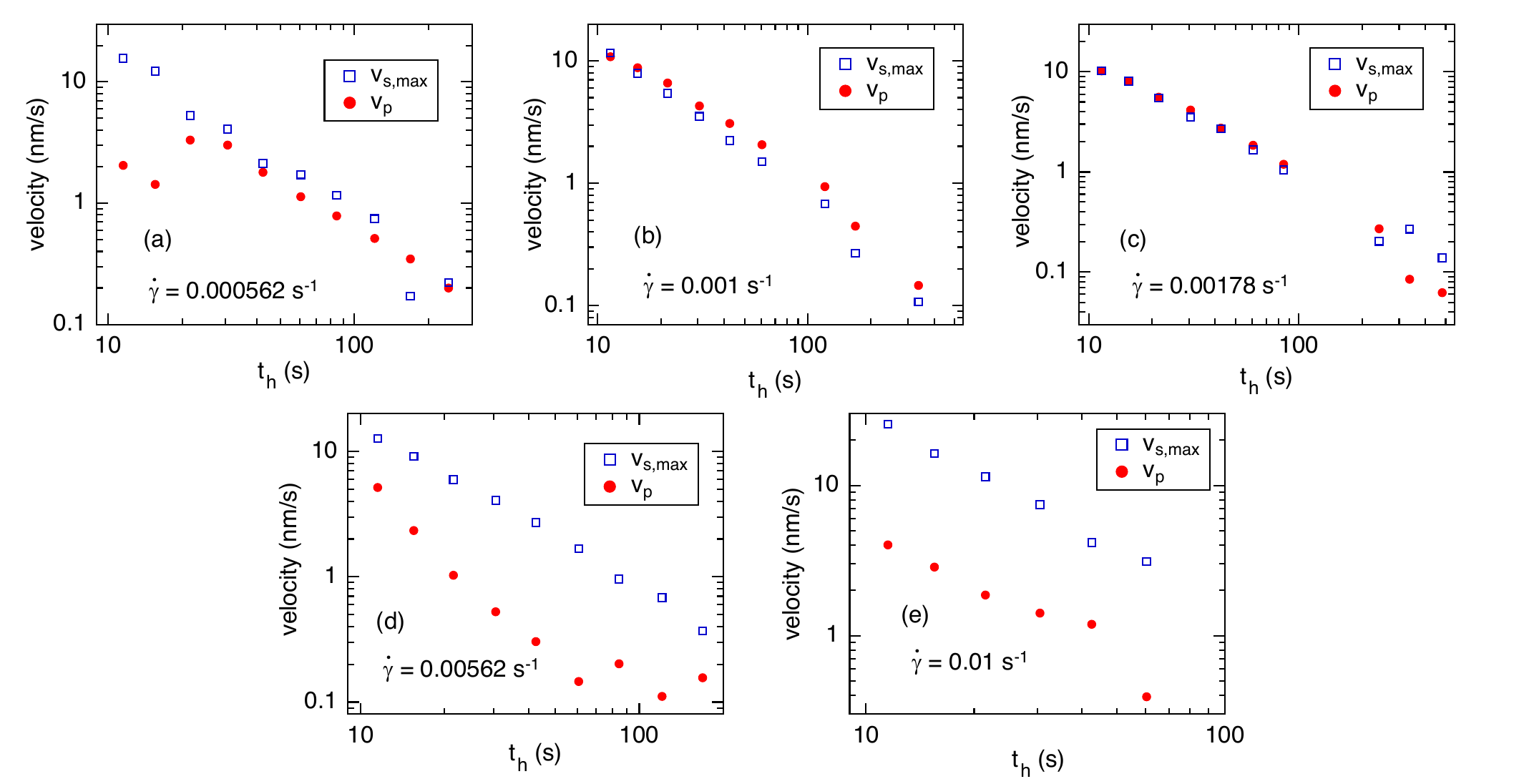}
\caption{Maximum velocity of the affine deformation (blue squares) and the plug flow velocity (red circles) as functions of hold time following cessation of shear flow to 300\% strain at shear rates of (a) 0.000562 s$^{-1}$, (b) 0.001 s$^{-1}$, (c) 0.00178 s$^{-1}$, (d) 0.00562 s$^{-1}$, and (e) 0.01 s$^{-1}$.}
\label{velocities}
\end{figure*}

Finally, Fig.~\ref{velocities} shows the maximum velocity of the affine strain and the plug flow velocity at different hold times following shear flow at the different initial shear rates.  Notably, at the lower initial shear rates, the two velocities are nearly the same at most hold times.

\end{document}